\documentclass[twocolumn,prl,superscriptaddress,amsmath,amssymb,nobibnotes]{revtex4-1}
\usepackage[latin9]{inputenc}
\usepackage{amsmath}
\usepackage{amssymb}
\usepackage{graphicx}

\makeatletter
\usepackage{color}
\usepackage[pdftex,dvipsnames]{xcolor}
\usepackage{upgreek}
\usepackage{float}
\usepackage{upgreek}
\usepackage{xargs}
\usepackage[colorinlistoftodos,prependcaption,textsize=tiny]{todonotes}

\usepackage{bibunits}
\defaultbibliography{Opto.bib}
\defaultbibliographystyle{unsrt}

\newcommandx{\unsure}[2][1=]{\todo[linecolor=red,backgroundcolor=red!25,bordercolor=red,#1]{#2}}
\newcommandx{\change}[2][1=]{\todo[linecolor=blue,backgroundcolor=blue!25,bordercolor=blue,#1]{#2}}
\newcommandx{\info}[2][1=]{\todo[linecolor=OliveGreen,backgroundcolor=OliveGreen!25,bordercolor=OliveGreen,#1]{#2}}
\newcommandx{\improvement}[2][1=]{\todo[linecolor=Plum,backgroundcolor=Plum!25,bordercolor=Plum,#1]{#2}}
\newcommandx{\thiswillnotshow}[2][1=]{\todo[disable,#1]{#2}}

\newcommand{\prlsec}[1]{\textit{#1.---}}

\makeatother

\begin{document}
\begin{bibunit}

\title{Quantum displacement sensing and cooling in 3D levitated cavity optomechanics}

\author{M. Toro\v{s}}
\affiliation{Department of Physics and Astronomy, University College London, Gower
Street, London WC1E 6BT, United Kingdom}

\author{T.S. Monteiro}
\email{t.monteiro@ucl.ac.uk}
\affiliation{Department of Physics and Astronomy, University College London, Gower
Street, London WC1E 6BT, United Kingdom}

\begin{abstract}
Ultra-high sensitivity detection of quantum-scale displacements in
cavity optomechanics  optimises the combined errors from measurement
back-action and  imprecisions from incoming quantum noises.
This sets the well-known Standard Quantum Limit (SQL).  Normal quantum
cavity optomechanics allows cooling and detection of a single
degree of freedom,  along the cavity axis. However, a recent
breakthrough that allows quantum ground-state cooling of  levitated
nanoparticles [Delic et al, arxiv:1911.04406], is uniquely 3D in character, with  coupling along the $x$,
$y$ and $z$ axes. We investigate current experiments  and show that the underlying behaviour is far from the
addition of independent 1D components and that ground-state cooling and sensing  analysis must consider- 
 to date neglected- 3D hybridisation effects. We characterise  the additional 3D spectral contributions and find  direct and indirect hybridising pathways can destructively interfere suppressing of 3D effects at certain  parameters  in order to approach, and possibly surpass, the SQL. We identify a  sympathetic cooling mechanism that can  enhance cooling of  weaker coupled modes, arising from optomechanically induced correlations.
\end{abstract}
\maketitle
The coupling of mechanical motion to the optical
mode of a cavity permits not only strong cooling but also ultra-sensitive
 displacement detection, and has led to advances ranging from quantum ground
state cooling of  mechanical oscillators \cite{Bowenbook,CavOptReview}
 to detection of gravitational waves by LIGO~\cite{LIGO}. Optomechanics employing
  levitated dielectric particles has recently also experienced rapid development \cite{LevReview2019,Yin2013}.
The unique potential of levitated cavity optomechanics in terms of
decoupling from environmental heating and decoherence, coupled with
the sensitivity of displacement sensing offered by optical cavities
was already recognised in 2010 \cite{RomeroIsart2010,Chang2010,Barker2010a}.
Actual experimental realisations represent a formidable technical
challenge: the levitated nanoparticle must be cooled from room temperatures,
and is initially millions of quanta above the quantum ground state. 

Most
initial proposals were for self-trapping set-ups \cite{Chang2010,Pender2012,Monteiro2013},
with trapping and cooling both provided by the cavity modes \cite{Kiesel2013},
but this failed to allow stable trapping  at high
vacuum \cite{Monteiro2013,Kiesel2013,Asenbaum2013}.
In order to overcome this roadblock, hybrid set-ups combining for
instance a tweezer and cavity traps \cite{RomeroIsart2010,Mestres2015};
or a hybrid electro-optical trap \cite{Millen2015,Fonseca2016}, or
a tweezer and near-field of a photonic crystal \cite{Magrini2018},
allowed some progress towards the ultimate goal of quantum ground
state cooling. 

This year, an important breakthrough was 
the realisation that the tweezer trapping light coherently scattered (CS) into an undriven
cavity offers major advantages~\cite{vuletic2001three,leibrandt2009cavity,hosseini2017cavity}:
the resulting optomechanical couplings  along every axis can be comparatively large even for
modest mean cavity photon numbers, minimising the deleterious
effects of photon scattering~\cite{Windey2018,Delic2018,GonzalezBallestero2019}.
As a result, quantum cooling of the centre of mass of a levitated nanoparticle  to phonon occupancies $n_x < 1$  along the $x$  axis (see Fig.1 for definition of axes) ) was recently reported \cite{Delic2019}.

 Here we investigate the 3D cooling and displacement sensing for CS systems. We obtain expressions for 3D  spectra  that reproduce  experimental features,  and yield excellent agreement with stochastic numerics using the  tweezer and cavity potentials without linearisation. We consider  direct intermode couplings overlooked previously and find  they introduce interference pathways that can (tunably) cancel  hybridisation between modes, without which the spectra and SQL analysis cannot in general be understood. 
While multi-mechanical-mode set-ups are not unusual in cavity optomechanics, typically those modes have widely differing quality factors or effective masses. In contrast, the  fully equivalent and strongly cooled modes here offer a new and unparalleled range of hybridisation and mutual back-action effects. 
In the experimental regimes of \cite{Delic2019}, we find that a strongly-cooledd $x$ mode is cooled to phonon occupancy $n_x \sim 1$, but  conclude that inclusion of  hybridisation effects is essential for reliable thermometry. Separately, a weakly-coupled $y$ mode experiences sympathetic cooling mechanism that lowers  $n_y$ significantly, due to optomechanical correlations, analogous to the ponderomotive squeezing mechanism, but between {\em mechanical} modes.

\begin{figure}[ht]
{\includegraphics[height=3.0in]{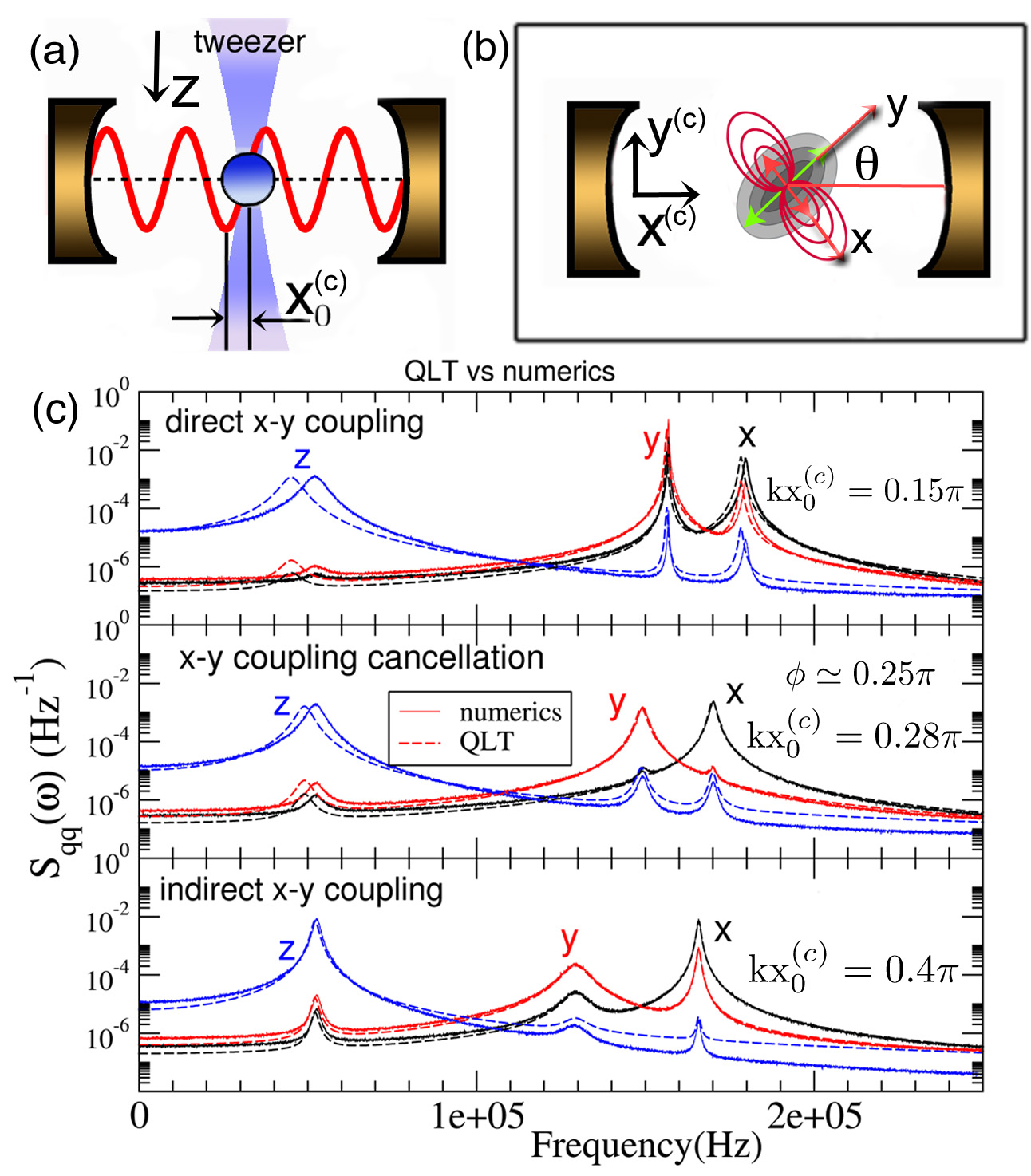}} \caption{ \textbf{(a)} Schematic of 3D cooling set-up in levitated optomechanics:
a nanoparticle held by a tweezer trap within a cavity. The cavity
is undriven, but is populated by photons coherently scattered from
the tweezer. The nanoparticle is placed at a point $\phi\simeq kx_{0}^{\text{(c)}}$
from the anti-node of the cavity field. Cooling and detection of the
centre of mass displacement in 3D along $x,y,z$ is possible. \textbf{(b)}
The pattern of coherent photon scattering (taken from \cite{Delic2018})
into the cavity depends on the tilt $\theta$ of the tweezer polarization
axis. \textbf{(c)} Compares displacement PSDs  using analytical expressions for the 3D theory
 (dashed lines) with stochastic numerics using the  tweezer
and cavity potentials (solid lines). The latter does not assume any values for the optomechanical
coupling strengths or equilibrium positions, and includes nonlinearities. $x$ (black), $y$ (red) and $z$ (blue).
Agreement between analytics and numerics is excellent. The high degree of hybridisation of   $x,y$ modes 
 results in a prominent double-peaked structure (even for $\Delta\gg\omega_{k}$ ) for (i) low $\phi$ (top
panel), because of direct coupling $g_{xy}$ and (ii) large $\phi\sim0.4\pi$
(bottom panel) because of indirect cavity mediated coupling $g_{x}g_{y}$. In contrast,
{\em suppression of hybridisation} is seen at  $\phi\approx\pi/4$  (middle panel), where destructive
interference between the direct and indirect pathways 
decouples the modes. 
Parameters in (a) similar to the experiment in \cite{Delic2018}:
input power $P_{\text{in}}=0.17$W, $\Delta=-300$ kHz; however sphere
radius $R_{0}=100$nm and finesse $\mathcal{F}=150,000$ are slightly
larger, with gas pressure $P=10^{-6}$ mbar, and  $\theta=0.2\pi$.}
\label{Fig1} 
\end{figure}

\prlsec{Displacement sensing} For a cavity mode $\hat{a}$, 
 displacement sensing will involve a measurement of
some quadrature of the optical field $\hat{Q}^{\Phi}=e^{-i\Phi}\hat{a}+e^{i\Phi}\hat{a}^{\dagger}$,
with coupling to a mechanical displacement $\hat{q}$, usually set
by the cavity axis, with coupling strength $g$ described by the well
known equation of linearised optomechanics: 
\begin{equation}
\hat{Q}^{\Phi}(\omega)=  \textrm{i}g \eta^{(\Phi)} \hat{q}(\omega)+\sqrt{\kappa}\tilde{Q}^{\Phi}_{\textrm{in}},\label{1DOpt}
\end{equation}
where $\tilde{Q}^{\Phi}_{\textrm{in}}$ represent measurement imprecision,
typically from incoming quantum photon shot-noise, while
$\kappa$ is the cavity linewidth. $\eta^{(\Phi)}\equiv \eta^{(\Phi)}(\omega)$ is the optical susceptibility, describing
the spectral shape of the cavity resonance. 
Understanding the Standard Quantum Limit (SQL) of displacement sensing in optomechanics usually
proceeds via analysis of errors in Eq.~\eqref{1DOpt} or related
forms.

In  3D, the measured optical quadrature in general  couples
to displacements $\hat{q}_{j}$ along all directions $j=x,y,z$: 
\begin{equation}
\hat{Q}^{\Phi}(\omega)=i\sum_{j} g_{j} \eta^{(\Phi_{j})} \hat{q}_{j}(\omega)+\sqrt{\kappa}\tilde{Q}^{\Phi}_{\textrm{in}},\label{3DOpt}
\end{equation}
where $\Phi_{j}\equiv\Phi$ for the normal optomechanical case where
displacement couples to the amplitude of the light, but $\Phi_{j}\equiv\Phi-\pi/2$
for the new scenario in the CS experiments \cite{Windey2018,Delic2018,GonzalezBallestero2019}
where it can couple to the optical phase quadrature. Above,  $\eta^{(\Phi)}=e^{-i\Phi}\chi(\omega,-\Delta,\kappa)-e^{i\Phi}\chi^{*}(-\omega,-\Delta,\kappa)$
where $\chi(\omega,-\Delta,\kappa)=[-\textrm{i}(\omega+\Delta)+\frac{\kappa}{2}]^{-1}$
and $\Delta$ is the detuning of the light from the cavity resonance.

In the well-known quantum linear theory (QLT) of cavity optomechanics \cite{Bowenbook,CavOptReview},  the 1D case is straightforward: the displacement
spectra are calculated from cavity amplified noise fluctuations $\tilde{\mathcal{D}}_{j}^{1D}$,
comprising thermal fluctuations of the mechanical modes in addition
to the fluctuations representing the back-action effect of the incoming
photon shot-noise. Neglecting certain normalisation terms (see \cite{SuppInfo}
for full-details) we have: 
\begin{equation}
\hat{q}_{j}(\omega)\equiv\tilde{\mathcal{D}}_{j}^{1D}\simeq\sqrt{\Gamma}\tilde{Q}_{j}^{therm}(\omega)+i\sqrt{\kappa}g_{j}\mu_{j}(\omega)\tilde{Q}^{(\Phi=0)}_{\textrm{in}}\label{1DDisplace}
\end{equation}
where $\Gamma$ is a mechanical damping, and $\mu_{j}(\omega)=\chi(\omega,\omega_{j},\Gamma)-\chi^{*}(-\omega,\omega_{j},\Gamma)$
is a mechanical susceptibility function that determines the back-action
spectrum generated by incoming quantum shot noise $\tilde{Q}^{(\Phi=0)}_{\textrm{in}}(\omega)$.
In the above 1D equations, the $\tilde{Q}_{j}^{therm}(\omega)$ might
represent the true signal we wish to measure, while the imprecision
and measurement back-action contributions in Eqs.~\eqref{1DOpt}
and \eqref{1DDisplace} represent measurement errors: minimising their
combined effect yields the well-known SQL \cite{Bowenbook,CavOptReview}.

With a simple adjustment to relate the intracavity field to the cavity
output field via input-output relations,
the corresponding PSD of the measured signal is used to estimate a
displacement spectrum $S_{\hat{Q}^{\Phi}\hat{Q}^{\Phi}}\simeq g^{2}|\eta^{(\Phi)}|^{2}S_{qq}^{1D}$
in the 1D case. A key question is whether one might straightforwardly
extend to the 3D displacement spectra by simply considering the sum
of the independent PSD contributions $S_{\hat{Q}^{\Phi}\hat{Q}^{\Phi}}\simeq\sum_{j=x,y,z}g_{j}^{2}|\eta^{(\Phi_{j})}|^{2}S_{q_{j}q_{j}}^{1D}$.
We show below that this is not the case. 
%We find that the most significant
%differences are new 3D back-action terms that redistribute energy
%between mechanical modes and add additional optical back-action; as
%we show, attaining the 3D SQL requires suppression of these terms.

\prlsec{3D Cavity optomechanics} As a first approximation to a 3D
system, one might simply replace, in Eq.~\eqref{3DOpt}, $\hat{q}_{j}(\omega)\rightarrow\tilde{\mathcal{D}}_{j}^{1D}$
and directly obtain the PSD for the homodyne spectrum, in other words
replace the displacement noises by their 1D equivalents. We note that
even in this straightforward case, the error analysis does not simply
yield a sum of the 1D PSDs $S_{q_{j}q_{j}}^{1D}$:
while the thermal contributions are uncorrelated and thus contribute
independently to the PSDs, the separate back-actions are all correlated
with each other and with the imprecision noises. This is important: even in the 1D case, correlations between
back-action and imprecision underlie well-known observed quantum spectral signatures such as sideband asymmetries and
optical (ponderomotive) squeezing. Correlations between optical back
action and imprecision noise also play an important role in LIGO displacement
sensing~\cite{Aggarwal2019}.

Our key findings is that we find additional, genuinely
3D, contributions and we can write the displacement noise spectrum
in the form: 
\begin{equation}
\hat{q}_{j}(\omega)\simeq\tilde{\mathcal{D}}_{j}^{1D}+\sum_{k\neq j}\mathcal{G}_{jk}^{3D}(\omega)\tilde{\mathcal{D}}_{k}^{1D}\label{3DNoise}
\end{equation}
from which we can obtain all PSDs analytically. Specifically, each
displacement, in addition to the usual 1D noises terms, receives contributions
from the 1D noises of the other two degrees of freedom, determined
by a 3D coupling function $\mathcal{G}_{jk}^{3D}(\omega)$ which we
can give in closed form and which quantifies the deviation from 1D
behaviour (numerical precision includes higher order correction
terms, see \cite{SuppInfo}, though
for clarity we discuss only the lowest order here).

To understand $\mathcal{G}_{jk}^{3D}(\omega)$, we revisit the quadratic forms of the
Hamiltonians  of linearised optomechanics,  obtained by considering small displacements
from an equilibrium point $(x_{0},y_{0},z_{0},{\bar{\alpha}})$ where
the mean photon number in the cavity is $n_{p}=|\bar{\alpha}|^{2}$.
Usually one writes  $\hat{H}/\hbar=\hat{h}^{(0)}+\sum_{j}\hat{h}_{j}^{\text{(int)}}$
where $\hat{h}^{(0)}=-\Delta\hat{a}^{\dagger}\hat{a}+\sum_{j}\omega_{j}\hat{b}_{j}^{\dagger}\hat{b}_{j}$,
$\hat{h}_{j}^{\text{(int)}}=g_{j}(\hat{a}^{\dagger}+\hat{a})\hat{q}_{j}$,
$\hat{q}_{j}=\hat{b}_{j}^{\dagger}+\hat{b}_{j}$, and we have $\Delta<0$
for a red-detuned cavity. 

However,  the full Hamiltonian
to quadratic order should be 
$\frac{\hat{H}}{\hbar}=\hat{h}^{(0)}+\sum_{j}\hat{h}_{j}^{\text{(int)}}+\sum_{j<k}g_{jk}\hat{q}_{j}\hat{q}_{k}$
where the last term on the right-hand side contains (previously neglected) direct coupling
terms of strength $g_{jk}$. These are distinct from nonlinear, position
squared coupling terms $g_{j}(\hat{a}^{\dagger}+\hat{a})\hat{q}_{j}^{2}$
which lead to observed sidebands at $2\omega_{j}$ in optically trapped
systems at higher temperatures \cite{Fonseca2016,Delic2018}.

In particular, starting from the Hamiltonian including $g_{jk}$ couplings,
where we have assumed the usual amplitude quadrature coupling, we
obtain: 
\begin{equation}
\mathcal{G}_{jk}^{3D}(\omega)=\frac{i\mu_{j}(\omega)}{M_{j}(\omega)}\left[i\eta^{(0)}(\omega)g_{j}g_{k}+g_{jk}\right].\label{3Dcouple}
\end{equation}
The prefactor, where $\mu_{j}(\omega)$ is the mechanical susceptibility
and $M_{j}=1+g_{j}^{2}\mu_{j}\eta^{(0)}$ is a function peaked around
one of the mechanical frequencies, i.e. $\omega\approx\pm\omega_{j}$.
However, it is the terms in the square brackets that are of most interest.
One can see they describe the interference between a direct, $\propto g_{jk}$,
and a cavity mediated, indirect coupling, $\propto g_{j}g_{k}$, between
any two displacements. In other words, suppressing or conversely,
enhancing 3D dynamics will involve either suppressing or correspondingly
enhancing the 3D coupling via destructive or constructive interference
of direct and indirect pathways near $\omega\approx\omega_{j}$.

\begin{figure}[!t]
{\includegraphics[width=3.in]{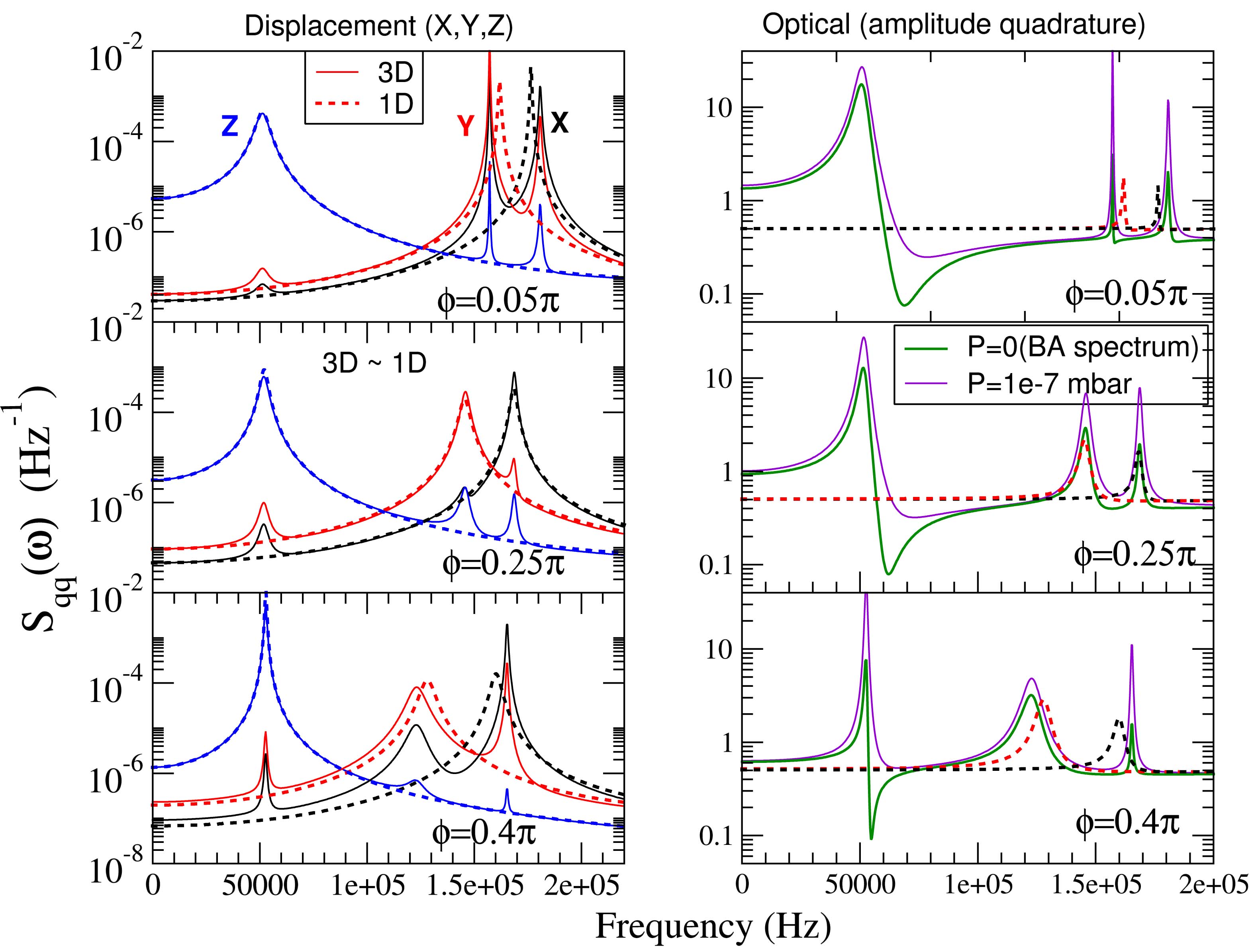}} \caption{\textbf{(a)} Comparison between full 3D QLT (solid lines) with the
equivalent 1D QLT (dotted lines) for the PSDs of the $x$, $y$, and
$z$ displacements (denoted by black, red, and blue colors, respectively).
While 3D QLT includes all optomechanical couplings $g_{x},g_{y},g_{z}$
and $g_{xy},g_{xz},g_{yz}$ for the full coupled problem, 1D QLT obtains
three independent PSDs $S_{q_{j}q_{j}}^{1D}$ with all couplings set
to zero except $g_{j}$. Parameters are similar to Fig.~\ref{Fig1},
with $\text{-}\Delta\gg\omega_{j}$, but pressure is set to $P=10^{-7}$
mbar for phonon occupancies near the quantum regime $n\sim1-3$. While
in general the $x$ and $y$ 3D PSDs are strongly perturbed (have
a double-peaked structure) we see that for $\phi\simeq\pi/4$ (middle
panels) they are very close to their 1D forms as there is destructive
interference between direct and indirect pathways. The $z$ mode contributes
only weakly as it is well separated in frequency. At lower (higher)
$\phi$ there are large differences between 1D and 3D PSDs due to
the direct (indirect) pathways as seen in the top (bottom) panels.
\textbf{(b)} For the optical output spectra (corresponding to homodyne
detection of the amplitude quadrature of the cavity output, violet
lines) the very large squeezing by the $z$ mode at $\phi\sim0$ lowers
the imprecision floor for the $x$ and $y$ PSDs. For comparison we
plot also the measurement back-action (BA) spectra (green curves)
obtained for $P\to0$. Dotted lines are the 1D BA equivalent and once
again, at $\phi=\pi/4$ these are also very close to the 3D form.}
\label{Fig2} 
\end{figure}

\prlsec{Tweezer-cavity setup} The above is quite generic to an arbitrary
3D optomechanics set-up. Here we apply this to the new experiments
pioneered in \cite{Windey2018,Delic2018} which involve levitating
a dielectric nanoparticle in a tweezer within a cavity. The tweezer
polarization and the cavity axis are tilted at an angle $\theta$
(see Figs.~\ref{Fig1}(a) and \ref{Fig1}(b)). The cavity in these
set-ups is undriven but is populated entirely by light coherently
scattered from the tweezer field and the particle moves under the
combined effect of the tweezer trapping field and the coherently scattered
light as explained in \cite{Windey2018,Delic2018}. We give the full
potential in \cite{SuppInfo}, but to a good approximation, the tweezer
represents a trapping Hamiltonian equivalent to $\hat{h}^{(0)}$,
while the interaction with the cavity mode yields the potential:
\begin{alignat}{1}
\frac{\hat{V}_{\text{int}}}{\hbar} & =-E_{d}\cos[\phi+k(\hat{x}\sin\theta+\hat{y}\cos\theta)]\hat{a}e^{-i\beta(\hat{z})}+\text{H.c.}\label{eq:Hint}
\end{alignat}
where $\phi\sim kx_{0}^{\text{(c)}}$, $x_{0}^{\text{(c)}}$ is the
displacement between the tweezer focus and an antinode of the cavity
(see Fig.~\ref{Fig1}(a)), $\beta(z)=kz-\arctan(z/z_{R})$, $z_{R}$
is the Rayleigh range, and $E_{d}$ is the coupling rate determined
by the particle polarisability and input power to the tweezer. Expanding
$\hat{V}_{\text{int}}$ to quadratic order provides the light-matter
couplings $g_{j}$, the matter-matter couplings $g_{kj}$ as well
as corrections to the mechanical frequencies and the equilibrium points
(see \cite{SuppInfo} for details).

The direct coupling has not previously been considered in the experimental
analysis \cite{Windey2018,Delic2018,GonzalezBallestero2019} but we
find they can be of great importance; one can show that $g_{xy}\simeq-g_{x}g_{y}\frac{2\text{Re}(\bar{\alpha})\cos{\phi}}{E_{d}\sin^{2}{\phi}}$,
while $g_{jz}\simeq g_{j}g_{z}\frac{2\text{Im}(\bar{\alpha})}{E_{d}\cos{\phi}}$
for $j=x,y$. Since $\bar{\alpha}\simeq\-iE_{d}\text{cos}(\phi)[\Delta+i\kappa/2]^{-1}$
we then readily find: 
\begin{equation}
g_{xy}\simeq g_{x}g_{y}\left[\frac{2\Delta\cot^{2}{\phi}}{\Delta^{2}+\frac{\kappa^2}{4}}\right],~g_{jz}\simeq g_{j}g_{z}\left[\frac{\kappa}{\Delta^{2}+\frac{\kappa^2}{4}}\right]\label{gjk}
\end{equation}
Thus depending on the positioning, $\Delta$ or $\kappa$, the direct
couplings contribution can be similar or exceed the cavity mediated
coupling.

In Fig.~\ref{Fig1}(c) we compare analytical, closed form PSDs we obtained
with 3D QLT and Eq.~\eqref{3DNoise}, with direct solutions of the
nonlinear Langevin equations of motion, using the tweezer and
cavity potential functions. In the latter, the $g_{j}$ and $g_{jk}$ are not  parameters but
rather simply emergent properties in the limit of low-amplitude displacements. 
The symmetrised analytical quantum spectra show excellent agreement with numerics in both quantum regimes as well as thermal (higher pressure regimes) provided the latter are cooled enough so that nonlinearities do not
generate additional peaks in the optical spectra \cite{Fonseca2016}. Furthermore, Fig.~\ref{Fig1}(c) also demonstrates the importance of the previously neglected $g_{kj}$ terms: in particular, leading to double peaked structures ($x-y$ hybridisation)
 for $\phi\simeq0$ where cavity mediated coupling $g_{x}g_{y}\simeq0$ terms are negligible, as well as
 $\phi\to\pi/2$, where $g_{xy}\to0$, but the cavity
mediated coupling from $g_{x}g_{y}$ are strong.

However the $\phi =\pi/4$ case is the most interesting and represents a key finding: here the $x$-$y$ hybridisation almost fully vanishes. Although both direct and indirect contributions are strong they interfere destructively.
We can show that $i\eta^{(0)}(\omega)\to\frac{-2\Delta}{(\kappa/2)^{2}+\Delta^{2}}$
if $\text{-}\Delta\gg\omega$ (and we are interested primarily in
the region $\omega\sim\omega_{j}$). Thus for large $-\Delta$, using
Eqs.~\eqref{3Dcouple} and \eqref{gjk}, we can readily show 
\begin{equation}
\mathcal{G}_{xy}^{3D}(\omega)\simeq g_{x}g_{y}\left[\frac{-2\Delta}{\Delta^{2}+(\kappa/2)^{2}}\right]\left[1-\cot^{2}\phi\right],\label{xycouple}
\end{equation}
and the $x,y$ coupling $\mathcal{G}_{xy}^{3D}(\omega)$ thus vanishes.

\begin{figure}[!t]
{\includegraphics[width=3.in]{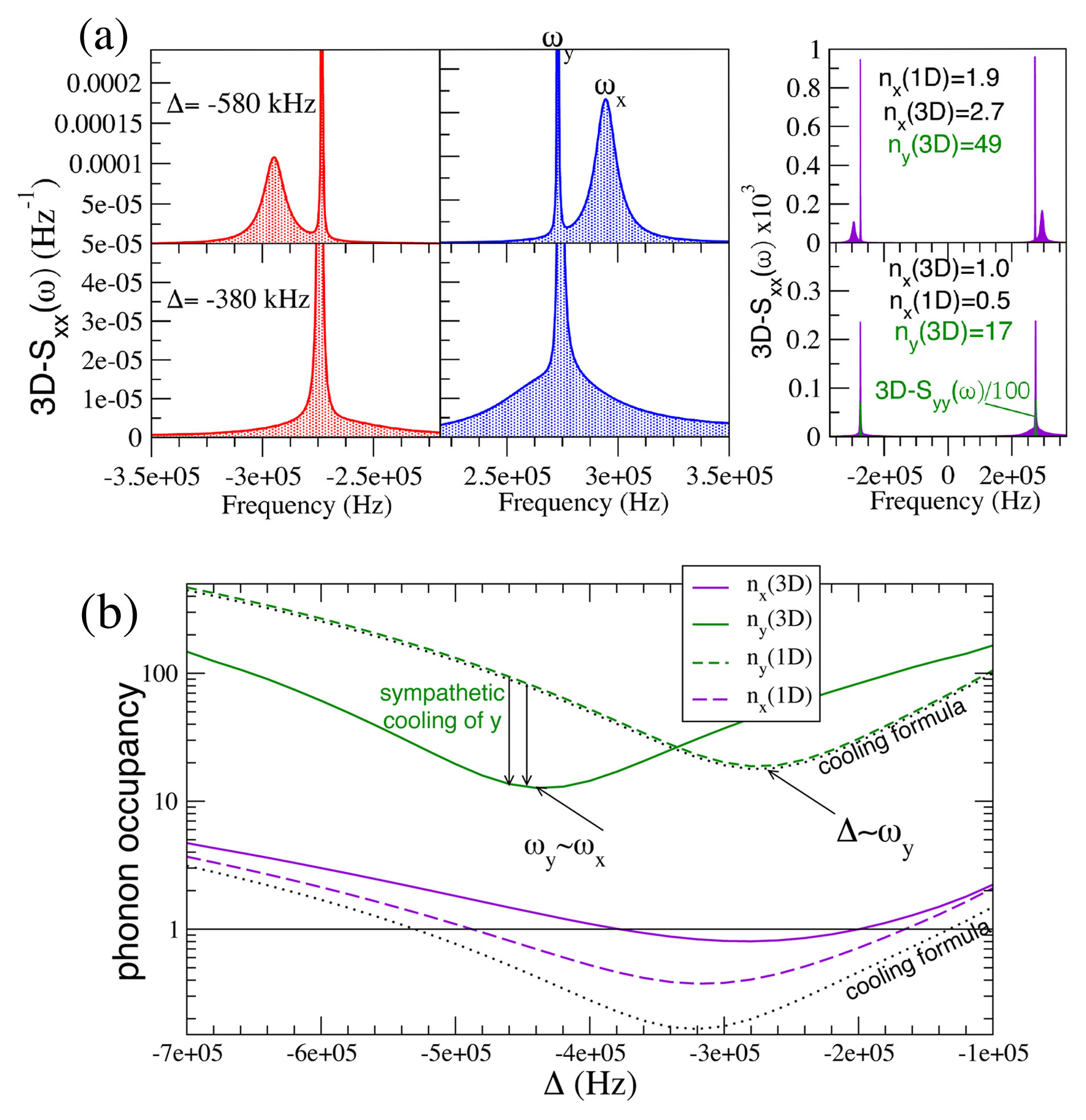}} \caption{Analysis of ground-state cooling experiments \cite{Delic2019}. 
  {\bf (a)}  Analytical $S_{xx}(\omega)$  PSDs reproduce well key experimental features of the blue and red sidebands. 
  $\phi=\pi/2$  so we conclude there is pure cavity mediated coupling $\propto g_xg_y$ as $g_{xy}\simeq 0$.  Also, $g_x\gg g_y$ since $\theta \simeq 0.47- 0.49 \pi$ {\bf (b)} Shows 1D vs 3D phonon occupancies $n_x,n_y$. 
The results in (a) and (b) expose several remarkable features. (i) The sharp  peak at $\omega \simeq \omega_y$, previously attributed to  $S_{yy}(\omega)$, in fact arises mainly from hybridisation,  hence the corresponding area should be considered in thermometry to validate measurements of  $n_x < 1$.  (ii) The narrow hybridisation peak shows very different asymmetry  from the main  broad $\omega\simeq \omega_x$ feature (see right panels of (a), showing both sidebands) so sideband asymmetry can only be measured from overall sideband area, not sideband heights. (iii) Surprisingly $n_y(\textrm{3D})$  can be almost an order of magnitude lower than $n_y(\textrm{1D})$,  due to a novel optomechanical sympathetic cooling effect; the $n_y$ minimum is displaced from the usual optomechanical cooling maximum at $\omega_y=-\Delta$. Dotted line plots in (b) results from the standard cooling formula of optomechanics that gives perfect agreement for 1D analytics, particularly  for the weak coupled $y$ mode.   $R_{0}=71.5$, $\mathcal{F}=73,000$, $P=10^{-6}$ mbar. Detailed analysis is in \cite{SuppInfo}.}
\label{Fig3} 
\end{figure}

We note $\phi\simeq\pi/4$, does not exactly correspond to  $kx_{0}^{\text{(}c)}=k\lambda/8$,  as there is an additional disturbance from co-trapping.  Double structures are seen in the experimental $x$ traces (see Fig.~\ref{Fig3}(c) of \cite{Delic2018}), directly detected via scattered light, which we tentatively attribute to hybridisation even for a particle placed $\lambda/8$ from the antinode.
The situation for the $\mathcal{G}_{jz}^{3D}(\omega)$ couplings is different as  the $z$  coupling is of the (non-standard for optomechanics)  form $g_{z}i(\hat{a}^{\dagger}-\hat{a})\hat{z}$. The cancellation of $z$ is partial, but
nevertheless, all 3D couplings are attenuated for $\text{-}\Delta\gg\omega_{j},\kappa$ (see \cite{SuppInfo} for details). Mixing with $z$ is weaker 
as typically, $\omega_z \ll \omega_x,\omega_y$.

The transition from  3D to a near decoupled 1D regime seen above  is further illustrated in Fig.~\ref{Fig2} (left panels) where we have compared
the PSDs obtained from 1D QLT (all $\mathcal{G}_{jk}^{3D}=0$) with
PSDs from the full 3D QLT.   In Fig.~\ref{Fig2} (right panels) we also look at the effect of
ponderomotive quantum squeezing both for thermal regimes
($P=10^{-7}$ mbar) as well as in the quantum back-action limit  ($P=0$). We see that as the $x,y,z$ contributions interfere, the strong squeezing by one mode ($z$) can lower the noise imprecision floor
for the other $x,y$ modes (upper right panel).

In Fig.~\ref{Fig3} we apply our theoretical analysis to the recent ground-state cooling experiments of \cite{Delic2019}
which are in the regime of $\phi=\pi/2$ (pure cavity-mediated coupling) and $\theta\simeq 0.47-0.49 \pi$ (hence $g_x\gg g_y$).  We reproduce the key experimental features but our analysis shows that the standard 1D analysis currently employed e.g. \cite{Delic2019} may not yield accurate thermometry and that hybridisation-related effects should be considered to establish  whether the precise $n_x<1$ threshold has been crossed (see \cite{SuppInfo} for details). 

\prlsec{Conclusions} We have shown that 3D optomechanical displacement
sensing can be far from a trivial sum of PSDs associated to the $\hat{x}$,
$\hat{\ensuremath{y}}$, and $\hat{z}$ degrees of freedom. Although
our work focusses on specifically on recent experiments on 3D  cooling
of levitated nanospheres, some of the conclusions are generic. We
show one may be able switch on and switch off some of the additional
3D effects and that these can give advantages in terms of exceeding
usual quantum back action limited occupancies for a given coordinate.
3D optomechanics opens the way to new forms of force and displacement
sensing, including sensing the direction as well as magnitude.

\prlsec{Acknowledgements}  We are extremely grateful to  Uro{\v{s}} Deli{\'{c}} for advice  and for sharing with us details 
of the experimental data. We acknowledge support from EPSRC grant EP/N031105/1.
%\bibliographystyle{aipauth4-1}  \bibliographystyle{unsrt}
%\bibliography{Opto}
\putbib
\end{bibunit}
 \clearpage
%\begin{widetext} 
\onecolumngrid
\begin{bibunit}
%%%%%%%%%% Prefix a "S" to all equations, figures, tables and reset the counter %%%%%%%%%%
\setcounter{equation}{0}
\setcounter{figure}{0}
\setcounter{table}{0}
\setcounter{section}{0}
\setcounter{page}{1}
\makeatletter
\renewcommand{\thesection}{S\arabic{section}}
\renewcommand{\theequation}{S\arabic{equation}}
\renewcommand{\thefigure}{S\arabic{figure}}
\renewcommand{\bibnumfmt}[1]{[S#1]}
\renewcommand{\citenumfont}[1]{S#1}
%%%%%%%%%% Prefix a "S" to all equations, figures, tables and reset the counter %%%%%%%%%%

 \section*{Supplementary Information}

Below we provide additional details of calculations in the main manuscript.  In Section I we discuss further our analysis of the recent experiment reporting ground state cooling of levitated nanoparticles. In Section II we discuss suppression of hybridisation with the $z$ motion.
In Section III we provide details of the derivation of our 3D QLT (Quantum Linear Theory) of optomechanics expressions. 
Finally in Section IV we review details of the potentials in the coherent scattering system and their linearisation in order to infer the optomechanical couplings $g_j$ as well as direct couplings $g_{jk}$ for $j,k=x,y,z$.

\section{Analysis of ground-state cooling experiments}

In this section we discuss the recent experiment reported in~\cite{Delic2019}
which employs the 3D coherent scattering setup discussed in the main text. The experiment places the particle at the node ($\phi=\pi/2$) and is thus in regime of pure indirect, cavity-mediated coupling, which differs significantly from the regimes where direct/indirect pathways compete and cancel. Nonetheless, there are other novel and important features. The
analysis confirms that the $\hat{x}$-motion is close to the ground
state and identifies  new effects in the $\hat{x}$ and $\hat{y}$ displacement spectra stemming
from hybridisation between the $\hat{x}$
to the $\hat{y}$ motions. In particular, we find  non-negligible
corrections to phonon occupancies in both modes.

\begin{figure}[ht]
{\includegraphics[width=6.6in]{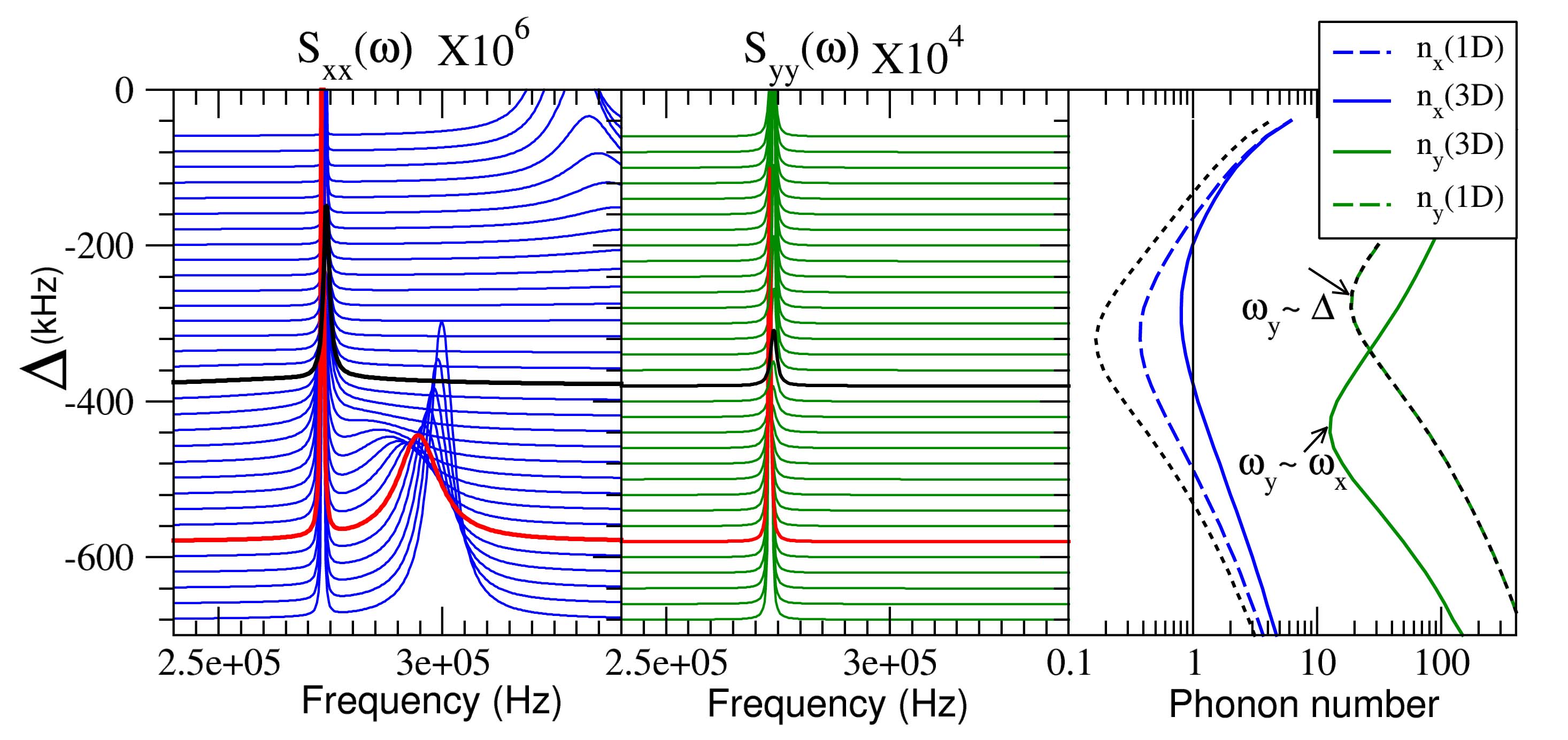}} \caption{ Shows PSDs corresponding to  $\hat{x}$ and $\hat{y}$ motions depicted using blue and green lines, respectively. The red (black) line correspond to
the detuning $\Delta=-2\pi\times580\text{kHz}$ ($\Delta=-2\pi\times380\text{kHz}$)
reported in~\cite{Delic2019}. The PSDs are in units of Hz$^{-1}$ but scaled as indicated for visibility. 
 (Left panel) PSDs for $\hat{x}$-motion at different detunings $\Delta$ showing two notable features:
(i) as the detuning approaches $-\Delta\sim\omega_{x}$, the $x$ motion
is optomechanically cooled to occupancies close to the ground state.
(ii) as $\Delta$ is lowered, since $g_x \gg g_y$, the optical spring effect  reduces $\omega_x$, but leaves $\omega_y$ unperturbed, resulting in a frequency degeneracy  $\omega_{x}\sim\omega_{y}$ that enhances
 hybridization effects and 3D heating/cooling channels. 
In particular, we note that $S_{xx}$, the $\hat{x}$ PSD, contains a sharp peak at $\omega \simeq \omega_y$
due to the hybridisation.
 (Middle panel) $\hat{y}$-motion PSD $S_{yy}(\omega)$.  The new 3D hybridisation affect cause significant
 cooling when the mechanical frequencies are degenerate (black line). 
 (Right panel) Phonon occupancies $n_{x}$ and $n_{y}$ as a function of detuning, where 1D (3D) indicates
a simplified one-dimensional (full three dimensional) analysis. We
note that the phonon occupancy for the $\hat{x}$-motion indicates
$n_{x}(\text{3D})>n_{x}(\text{1D})$.
The $\hat{y}$-motion is cooled most effectively at the hybridization
point, $\omega_{y}\sim\omega_{x}$. This latter effect is in 
contrast with the behaviour expected from a simplified 1D analysis,
where cooling is most effective at $-\Delta=\omega_{y}$. Black dashed lines denote results from standard optomechanics cooling formula (see Eq.(\ref{eq:change1})). We take $\theta=0.47 \pi$. } 
\label{FigS1} 
\end{figure}

We have the following $x$-$y$ hybridisation coupling strengths (see also Sec.~\ref{sec:-3D-levitated} here):
 \begin{eqnarray}
\mathcal{G}_{xy}^{3D}(\omega)  =\frac{i\mu_{x}(\omega)}{M_{x}(\omega)}\left[i\eta^{(0)}(\omega)g_{x}g_{y}+g_{xy}\right] \  \ \textrm{and} \ \ 
\mathcal{G}_{yx}^{3D}(\omega)  =\frac{i\mu_{y}(\omega)}{M_{y}(\omega)}\left[i\eta^{(0)}(\omega)g_{x}g_{y}+g_{xy}\right].
\end{eqnarray}
However, as the nanoparticle is located at a cavity node,  $\phi=\pi/2$, they involve only the indirect cavity-field mediated coupling terms $\propto  \eta^{(0)}(\omega)g_{x}g_{y}$ since the direct coupling coefficients vanish, i.e. $g_{xy}=0$. Hence the $x$-$y$ hybridisation couplings in the case of pure cavity-mediated interactions reduce to:
\begin{equation}
\mathcal{G}_{jk}^{\text{3D}}=-\frac{\mu_{j}\eta_{0}g_{j}g_{k}}{1+g_{j}^{2}\mu_{j}\eta_{0}},\label{eq:Gjk}
\end{equation}
where $j$, $k$ denote the indices $x$ and $y$.  Interestingly, although in this configuration the cavity contains very few photons (only components at the Stokes/anti-Stokes frequencies), the indirect couplings  still play a very important role. 

In addition, the tweezer tilt-angle is set to values $\theta \approx \pi/2$, thus $g_y \ll g_x$, so one expects strong cooling exclusively along the $x$ direction.  However, the heterodyne detected PSDs showed prominent peaks at $\omega \simeq \omega_y$ (shifted by the reference oscillator) thus one infers that $g_y \neq 0$ so $\theta\neq \pi/2$.  Allowing for an uncertainty  in the tweezer tilt of a few degrees, we have thus assumed $\theta=(0.47-0.49)\pi$ to be consistent
with the observations. For $\theta=0.47\times\pi$, we obtain
 $g_{x}\approx 2\pi\times80\text{kHz}$ and $g_{y}\approx 2\pi\times8\text{kHz}$, thus $g_x \approx 10 g_y$.

A further detail of the observed data motivates a very small ($10\%$) adjustment of the tweezer waist dimensions.  
Fig.~\ref{FigS1} shows an optical-spring induced frequency degeneracy between the $x$ and $y$ modes at 
$\Delta \approx -400$ kHz, which is a feature of the experiments. In order to get agreement in the $x$, $y$ frequencies as well as the frequency degeneracy, the tweezer waist values $w_x=0.66\mu$m and $w_y=0.77\mu$m in \cite{Delic2019} were reduced slightly to $w_x=0.600\mu$m and $w_y=0.705\mu$m, which is consistent with inherent experimental uncertainties in the tweezer geometry.

Fig.~\ref{FigS1} illustrates key features of the experimental regime in \cite{Delic2019}, including the optical-spring induced degeneracy, the cooling dynamics, and the hybridisation. This is the scenario we now analyse.
The $z$ motion has a frequency $\omega_{z}\ll\omega_{x},\omega_{y}$, and can thus be neglected in the simplified analysis below (but is included in the numerics). In this regime the $x$ and $y$ mechanical motions form a system
of coupled equations, which in frequency space take the form:
\begin{alignat}{1}
\hat{x} & ={\tilde{D}}_{x}^{\text{1D}}+\mathcal{G}_{xy}^{\text{3D}}\hat{y},\label{eq:xe}\\
\hat{y} & ={\tilde{D}}_{y}^{\text{1D}}+\mathcal{G}_{yx}^{\text{3D}}\hat{x}.\label{eq:ye}
\end{alignat}
The terms ${\tilde{D}}_{x}^{\text{1D}}$ and $\tilde{\mathcal{D}}_{y}^{\text{1D}}$
denote the optical and mechanical noises which would be present already
in a one-dimensional analysis, and the hybridization couplings $\mathcal{G}^{\text{3D}}_{jk}$ are given in Eq.~\eqref{eq:Gjk}.

\subsection{Analysis of the $y$ motion}
Substituting Eq.(\ref{eq:xe}) into Eq.(\ref{eq:ye}) we find:
\begin{equation}
\hat{y} \ ={\tilde{D}}_{y}^{\text{1D}}+\mathcal{G}_{yx}^{\text{3D}}[\mathcal{\tilde{D}}_{x}^{\text{1D}}+\mathcal{G}_{xy}^{\text{3D}}\hat{y}],
\label{eq:corr}
\end{equation}
showing that the optomechanics introduces correlations between the $x$ and $y$ motions although the corresponding thermal noise fields are uncorrelated.  As we are operating relatively far from the backaction limit, we neglect
in the first instance the optical noises and hence the optically induced correlations between $\tilde{\mathcal{D}}_{x}^{\text{1D}}$ and $\tilde{\mathcal{D}}_{y}^{\text{1D}}$. We note however that the above optomechanically induced correlations between the $x$ and $y$ modes are somewhat analogous to the well-studies correlations between optical and mechanical modes induced by optomechanical backaction.

\begin{figure}[ht]
{\includegraphics[width=4in]{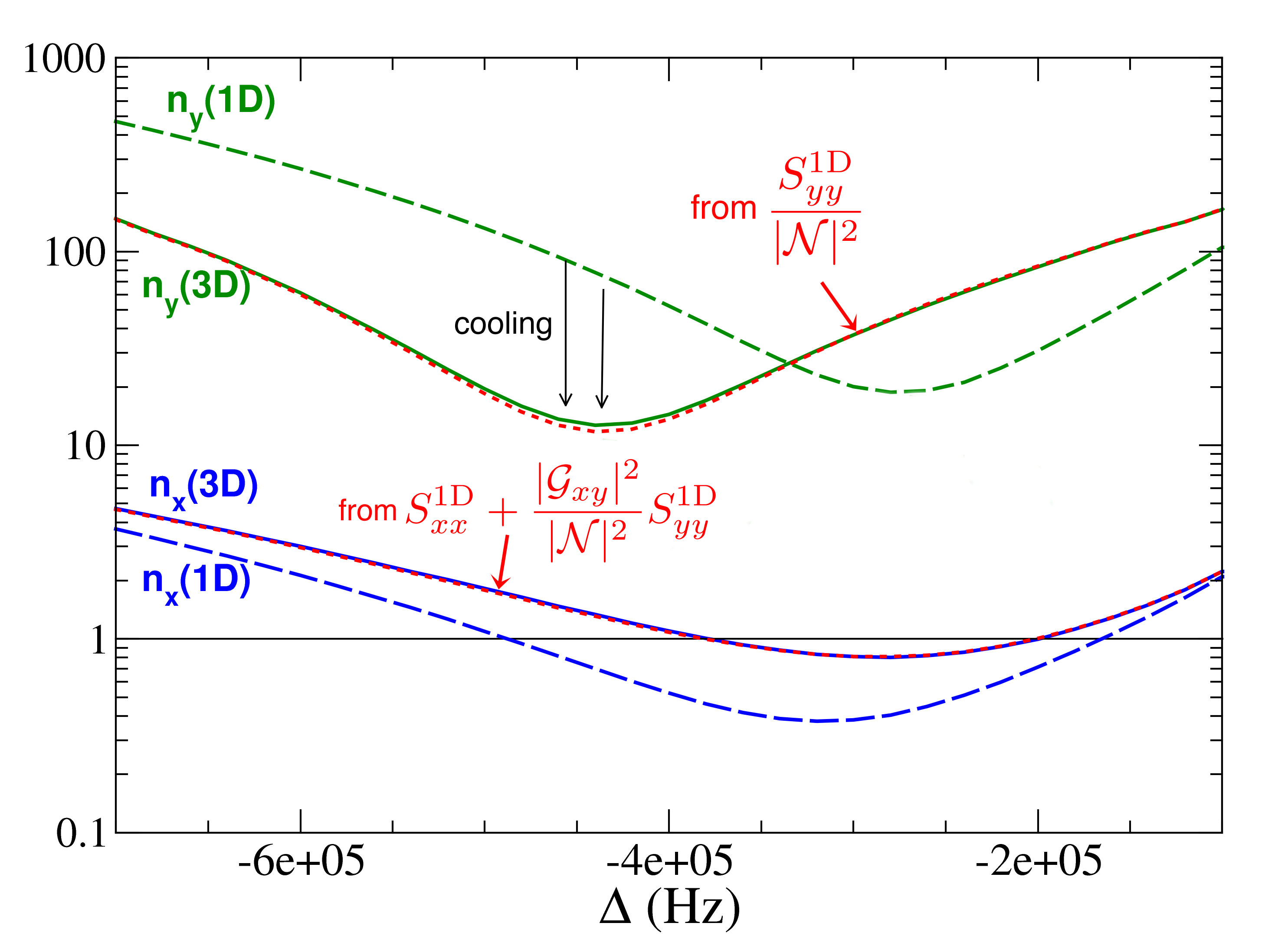}} \caption{Phonon occupancies $n_{x}$ and $n_{y}$ as a function of detuning 
showing that the 3D PSDs may be accurately estimated by a simple model that rescales the 1D PSDs 
 (Eq.(\ref{eq:Syy3D}) for $y$ and Eq.(\ref{eq:Sxx3D}); results showing the rescaled PSDs (in red) are in excellent agreement with the full 3D expressions.} 
\label{FigS2} 
\end{figure}

As $x$ is strongly cooled, we can in this case neglect the ${\tilde{D}}_{x}^{\text{1D}}$ term. Hence,
\begin{equation}
\hat{y} \simeq (1-\mathcal{G}_{yx}^{\text{3D}}\mathcal{G}_{xy}^{\text{3D}})^{-1} {\tilde{D}}_{y}^{\text{1D}} = 
\mathcal{N}^{-1}(\omega) {\tilde{D}}_{y}^{\text{1D}},
\label{eq:NDy}
\end{equation}
and thus we arrive at an an approximate expression for the PSD of $\hat{y}$:
\begin{equation}
S_{yy}^{\text{3D}}\simeq\frac{S_{yy}^{\text{1D}}}{\vert\mathcal{N}(\omega)\vert^{2}}.\label{eq:Syy3D}
\end{equation}
Fig.~\ref{FigS2} compares the above $\mathcal{N}$-rescaled PSD with the full analytical expressions, showing that the rescaling of the 1D sideband accurately accounts for the differences between the 3D and 1D PSDs including the relative heating and cooling.

In summary, around the frequency-degeneracy, there is strong (about factor 7) cooling of the $y$ motion due to the $x$-$y$ correlations and the backaction of $y$ on $x$, i.e. the $y$ mode is, via the cavity, coupled to $x$, and in turn the $x$ mode, because of this cavity-mediated coupling, acquires a component correlated with the $y$ thermal noises. We identify this as a new mechanism for ``sympathetic  cooling'' of the $y$ mode, due entirely to the strongly coupled (and strongly cooled) $x$ mode.

\subsection{Analysis of the $x$ motion}

The $x$ motion can be analysed in similar manner, by substituting Eq.(\ref{eq:ye}) into Eq.(\ref{eq:xe}) which readily gives 
\begin{equation}
\hat{x}=\mathcal{N}^{-1}(\omega)\left[\tilde{\mathcal{D}}_{x}^{\text{1D}}+\mathcal{G}_{xy}^{\text{3D}}\tilde{\mathcal{D}}_{y}^{\text{1D}}\right].\label{eq:xs2d}
\end{equation}
Analogously, we find the PSD:
\begin{equation}
S_{xx}^{\text{3D}} \simeq  S_{xx}^{\text{1D}}+ \frac{|\mathcal{G}_{xy}^{\text{3D}}|^2}{\vert \mathcal{N}\vert^{2}} 
S_{yy}^{\text{1D}}     \simeq  S_{xx}^{\text{1D}}+ |\mathcal{G}_{xy}^{\text{3D}}|^2  S_{yy}^{\text{3D}},   \label{eq:Sxx3D}
\end{equation}
where we have made the further approximation, based on inspection of the form of $\mathcal{N}(\omega)$, that
$S_{xx}^{\text{1D}} \simeq S_{xx}^{\text{1D}} |\mathcal{N}|^{-2}$ ; in other words, the backaction of highly cooled $x$ motion on the PSD of $x$, arising from its coupling to $y$,  is relatively unimportant.
The important difference between the  $S_{xx}^{\text{3D}}$ and $S_{xx}^{\text{1D}}$ arises from the 
second term in Eq.(\ref{eq:Sxx3D}). This latter term is not an interference term, but an additive term, which always results in additional heating, and it provides the sharply peaked feature around $\omega \simeq \omega_y$.

This feature has previously been neglected, but its contribution to the sideband area should be included for accurate thermometry.

We note that the $x$ sideband is strongly affected by the optomechanical spring effect. It is straightforward to adapt 
the usual analysis for this $x-y$ coupled case. One obtains the usual self-energy ~\cite{marquardt2007quantum}
and (see Sec.IV for more details):
\begin{equation}
\Sigma_{x}\equiv\frac{g_{x}^{2}\eta^{(0)}}{1+g_{y}^{2}\mu_{y}\eta^{(0)}}.
\end{equation}
from whence we  find the change of the damping, $\delta\Gamma$, and
the shift of frequency that represents the optical spring effect, $\delta\omega$, using the following expressions~\cite{marquardt2007quantum}:
\begin{equation}
\delta\Gamma_{j}=\frac{\text{Im}(\Sigma_{j}(\omega_{\text{j}}))}{\omega_{j}},\qquad\delta\omega_{j}=\frac{\text{Re}(\Sigma_{j}(\omega_{\text{j}}))}{2\omega_{\text{j}}},\label{eq:change1}
\end{equation}
where $\omega_{\text{j}}$ denotes the mechanical frequency.  Note the $g_y^2$ correction in the denominator 
of the self-energy; setting this to zero yields the standard 1D optomechanical cooling formula. 

 In Fig.~\ref{FigS1} we compared phonon occupancies (black dashed lines, right panel) obtained in this way $n_x\simeq  n_B \gamma_g/\Gamma_x$
where the thermal bath occupancy $n_B=kT/(\hbar \omega_x)$ for $T=300$K.
 We note that the effect of this $g_y^2$ correction is  small and the significant effects in heating of $x$ arise rather from the hybridisation correction (the last term in Eq.(\ref{eq:Sxx3D})). 

\subsection{Analysis of heterodyne-detected spectra: thermometry and sideband asymmetry}

In \cite{Delic2019} the area under heterodyne-detected sidebands was evaluated to estimate phonon occupancies. However, the sharp peak at $\omega \sim \omega_y$ was excluded. A 3D analysis including hybridisation indicates this is likely to underestimate the area and hence the final phonon occupancy. It is interesting to estimate what proportion of the peak is due to $y$ motion (and hence should be discounted when estimating $n_x$) and what proportion is due to hybridisation and thus contributes to the calculation of the energy in the $x$ motion.

Heterodyne detection will detect $y$ motion with amplitude  $\propto g_y^2 |\eta^{(0)}(\omega)|^2 S^{\text{3D}}_{yy}(\omega)$, (where  frequencies are shifted by the appropriate reference oscillator). 
We have shown that the 3D PSDs may be accurately estimated by a simple model employing rescaled PSDs
 (Eq.(\ref{eq:Syy3D}) for $y$ and Eq.(\ref{eq:Sxx3D}) for $x$. In turn, the hybridisation component in the heterodyne spectra
is given by the second term in Eq.(\ref{eq:Sxx3D}) but its heterodyne detection amplitude scales with $g_x^2|\eta^{(0)}(\omega)|^2$.
 
 The $y:x$ component ratios may be estimated $1 :R$ where 
 
\begin{equation}
 R= \frac{g^2_x}{g^2_y}   |\mathcal{G}_{xy}^{\text{3D}}(\omega \simeq \omega_y)|^2 
 \end{equation}
 
 It is straightforward to plot the function $|\mathcal{G}_{xy}^{\text{3D}}(\omega )|^2 $  and given $\frac{g^2_x}{g^2_y}\simeq 100$, we obtain $R=1.8$ at $\Delta=-580$ kHz.
 In the hybridisation region $\Delta=-380-420$ kHz, we find  $R\approx 3-5$  thus $\approx 70-90\%$ of the sharp peak
 is due to hybridisation and hence contributes to $n_x$ thermometry. 
 For the strongest cooling data, however at $\Delta=-300$ kHz, we find $R=1.1$ thus in the strongest $x$ cooling region, only half the $\omega\simeq \omega_y$ peak is due to hybridisation.  
 
 From the above analysis, we can see that the sharp peak to a good approximation carries the asymmetry of the $y$ motion. If one eliminates asymmetry introduced by the cavity susceptibility function $|\eta^{(0)}(\omega)|^2$ , the underlying asymmetry of the sharp $\omega \simeq \omega_y$ peak  is $n_y(\text{3D})+1 :n_y(\text{3D})$.   In contrast, the asymmetry of the broad feature is closer to $n_x(\text{1D})+1: n_x(\text{1D})$.
 
 This is in sharp contrast to the usual scenario in optomechanics where the red and blue sidebands have exactly the same shape but are simply rescaled by a factor $n/(n+1)$ where $n$ is the appropriate occupancy. Here, the unusual hybridisation means that the full area of the sidebands including the $y$ peak must be considered in order to estimate
 $n_x(\text{3D})$ from sideband asymmetry.

\section{Suppression of $z$ hybridisation}
In the main text  we found that  a  remarkable transition from  3D to a near decoupled 1D regime
occurs  for $-\Delta \gg \omega_{x,y}$ and $\phi=\pi/4$,  in between the 3D 
 (direct coupled, $\phi \simeq 0$) and 3D (indirect, cavity mediated. $\phi \simeq \pi/2$) regimes.
 This results from the cancellation between the direct $g_{xy}$ coupling and the cavity mediated
 $g_xg_y \eta^{(0)}$ terms; and underlying reason for this surprising near exact cancellation is that
 $g_{xy}\propto\text{Re}(\bar{\alpha})$ where $\bar{\alpha}$ is the mean cavity field, which follows the
cavity resonance, that in turn determines the form of $\eta^{(0)}$. 
 
However, the situation for the $\mathcal{G}_{jz}^{3D}(\omega)$ couplings is similar but
more involved ($j\neq z$) so the destructive cancellation is less complete.
 A peculiarity of the system is that the $z$  coupling is of the form $g_{z}i(\hat{a}^{\dagger}-\hat{a})\hat{z}$, i.e. the
displacement couples to the {\em momentum} quadrature of the cavity.
In this case, $\mathcal{G}_{jz}^{3D}(\omega)=\frac{i\mu_{j}(\omega)}{M_{j}(\omega)}\left[-i\eta^{(\pi/2)}(\omega)g_{j}g_{z}+g_{jz}\right]$,
but $\mathcal{G}_{zj}^{3D}(\omega)=\frac{i\mu_{z}(\omega)}{M_{z}(\omega)}\left[i\eta^{(\pi/2)}(\omega)g_{z}g_{j}+g_{jz})\right]$.
In other words, $\mathcal{G}_{jz}^{3D}\neq\mathcal{G}_{zj}^{3D}$
and both couplings cannot be suppressed simultaneously. In any case,
using the equation from the main text:
\begin{equation}
g_{xy}\simeq g_{x}g_{y}\left[\frac{2\Delta\cot^{2}{\phi}}{\Delta^{2}+\frac{\kappa^2}{4}}\right],~g_{jz}\simeq g_{j}g_{z}\left[\frac{\kappa}{\Delta^{2}+\frac{\kappa^2}{4}}\right]
\end{equation}
 and for large values of $\text{-}\Delta$
where $i\eta^{(\pi/2)}\to\frac{\kappa-2i\omega}{\Delta^{2}+(\kappa/2)^{2}}$,
we find that $\mathcal{G}_{jz}^{3D}(\omega)\propto\frac{2g_{j}g_{z}i\omega}{\Delta^{2}+(\kappa/2)^{2}}$
and $\mathcal{G}_{zj}^{3D}(\omega)\propto\frac{2g_{j}g_{z}[\kappa+i\omega]}{\Delta^{2}+(\kappa/2)^{2}}$.

Thus even where there is destructive interference, only the real part
of $\mathcal{G}_{jz}^{3D}(\omega)$ is fully cancelled.
Nevertheless, all 3D couplings are attenuated for $\text{-}\Delta\gg\omega_{j},\kappa$. 
Further, since fortunately  since $\omega_z \ll \omega_{x,y}$, hybridisation between $z$ and the other two
modes  is generally weaker than between $x$ and $y$ which are close in frequency. Thus it is possible to 
tune quite strongly into the decoupled 1D regime.

\section{Quantum Linear Theory (QLT)\label{sec:QLT}}

\subsection{Standard optomechanics QLT\label{soQLT} }

In this section we briefly review the framework of quantum linear
theory (QLT) of optomechanics. Optically levitated systems \cite{Millen2016,Aranas2017}
generally involve multiple optical and mechanical modes. Such multi-mode
systems ($N$ optical and $M$ mechanical degrees of freedom) are
typically described by the well-studied linearised Hamiltonian~\cite{Bowenbook}:
\begin{equation}
{\hat{H}}=\sum_{l=1}^{l=N}-\Delta_{l}{\hat{a}}_{l}^{\dagger}{\hat{a}}_{l}+\sum_{k=1}^{k=M}\omega_{k}{\hat{b}}_{k}^{\dagger}{\hat{b}}_{k}-\sum_{k,l}g_{k}^{(l)}({\hat{a}}_{l}^{\dagger}+{\hat{a}}_{l})({\hat{b}}_{k}^{\dagger}+{\hat{b}}_{k})\label{Hlin}
\end{equation}
where $\hat{a}_{l}$ $(\hat{a}_{l}^{\dagger})$ is the annihilation
(creation) operator for optical mode $l$, and $\hat{b}_{k}$ $(\hat{b}_{k}^{\dagger})$
for mechanical mode $k$. $\Delta_{l}$ is the detuning between the
input laser and the cavity mode $l$, while $\omega_{k}$ is the natural
frequency of the mechanical oscillator, and $g_{k}^{(l)}$ is the
light-enhanced coupling strength between an optical mode and a mechanical
mode. For simplicity, dissipation is characterised by a single optical
damping rate, $\kappa$, and a single mechanical damping rate, $\Gamma$
(though more complex scenarios, for example with multiple mirror losses,
can be easily incorporated).

A set of $2(N+M)$ quantum Langevin equations of motion are obtained
from Eq.~\eqref{Hlin} by adding input noises. For example, for the
single mode $N=M=1$ case, where all $g_{k}^{(l)}\equiv g$, we have:
\begin{equation}
\begin{pmatrix}\dot{\hat{a}}(t)\\
\dot{\hat{a}}^{\dagger}(t)\\
\dot{\hat{b}}(t)\\
\dot{\hat{b}}^{\dagger}(t)
\end{pmatrix}=\begin{pmatrix}\textrm{i}\Delta-\frac{\kappa}{2} & 0 & \textrm{i}g & \textrm{i}g\\
0 & -\textrm{i}\Delta-\frac{\kappa}{2} & -\textrm{i}g & -\textrm{i}g\\
\textrm{i}g & \textrm{i}g & -\textrm{i}\omega-\frac{\Gamma}{2} & 0\\
-\textrm{i}g & -\textrm{i}g & 0 & \textrm{i}\omega-\frac{\Gamma}{2}
\end{pmatrix}\begin{pmatrix}\hat{a}(t)\\
\hat{a}^{\dagger}(t)\\
\hat{b}(t)\\
\hat{b}^{\dagger}(t)
\end{pmatrix}+\begin{pmatrix}\sqrt{\kappa}\hat{a}_{\textrm{in}}(t)\\
\sqrt{\kappa}\hat{a}_{\textrm{in}}^{\dagger}(t)\\
\sqrt{\Gamma}\hat{b}_{\textrm{in}}(t)\\
\sqrt{\Gamma}\hat{b}_{\textrm{in}}^{\dagger}(t)
\end{pmatrix},\label{matrix-standard}
\end{equation}
where $\hat{a}_{\textrm{in}}$ ($\hat{b}_{\textrm{in}}$) is the optical
(mechanical) input noise. The above equation even for arbitrary numbers
of modes can be cast in matrix form: 
\begin{equation}
\dot{\mathbf{c}}(t)=\mathbf{A}\mathbf{c}(t)+\mathbf{c}_{\mathrm{in}}(t),\label{QLE}
\end{equation}
where the vector $\mathbf{c}=\begin{pmatrix}\hat{a}_{1} & \hat{a}_{1}^{\dagger}...\hat{a}_{N} & \hat{a}_{N}^{\dagger} & \hat{b}_{1} & \hat{b}_{1}^{\dagger}...\hat{b}_{M} & \hat{b}_{M}^{\dagger}\end{pmatrix}^{\mathsf{T}}$,
the matrix $\mathbf{A}$ contains the frequencies of the problem,
and $\mathbf{c}_{{\rm {in}}}$ are Gaussian input noises (incoming
quantum shot noise in the ideal case in the optical modes and thermal
noise for the mechanical noises).

Multi-mode theoretical PSDs are efficiently computed using a the Linear
Amplifier Model \cite{Botter2010}. For the LAM, the first step involves
transforming the equations of motion into frequency space. The coupled
equations are then manipulated analytically (or even numerically if
unavoidable) to recast the matrix equation of the equations of motion
in the form: 
\begin{equation}
\mathbf{c}(\omega)=\mathbf{T}\mathbf{c}_{\textrm{in}}(\omega),\label{LAM}
\end{equation}
where $\mathbf{T}(\omega)=\left(-\textrm{i}\omega\mathbf{I}-\mathbf{A}\right)^{-1}$
and $\mathbf{I}$ is the identity. $\mathbf{T}$ is a transformation
matrix that characterises the transduction of the input noises into
the mechanical and optical field fluctuations, somewhat analogous
to the effect of a linear amplifier. The linear amplifier model is
very powerful as one may in principle obtain the vector of all PSDs
of all modes in one go: 
\begin{equation}
S_{\mathbf{c}\mathbf{c}^{\dagger}}(\omega)=\mathbf{T}(\omega)\mathbf{N}\mathbf{T}^{\dagger}(\omega),\label{Scc}
\end{equation}
where 
\begin{equation}
\left<\mathbf{c}_{\textrm{in}}(\omega)[\mathbf{c}_{\textrm{in}}(\omega)]^{\dagger}\right>=\mathbf{N},\label{kronecker}
\end{equation}
and $\mathbf{N}$ is a diagonal matrix of elements: 
\begin{equation}
\mathbf{N}=\text{diag}\begin{pmatrix}\gamma_{1}(\bar{n}_{1}+1) & \gamma_{1}\bar{n}_{1} & \cdots & \gamma_{n}(\bar{n}_{n}+1) & \gamma_{n}\bar{n}_{n}\end{pmatrix}.\label{Nmatrix}
\end{equation}
The $n_{k}$ represent the occupancy of the respective baths, thus
$n_{k}=0$ for quantum shot noise in the optical modes but $n_{k}\simeq kT/\hbar\omega_{k}$
for thermally occupied phonon modes. Typically, one can set $\gamma_{k}\equiv\kappa$
for optical modes and $\gamma_{k}\equiv\Gamma$ for the mechanical
modes. For levitated systems there is no cryogenic cooling and $T=300$K.

The solutions $\hat{a}_{l}(\omega)$ of the optical field denote here
the intra-cavity field, while the actual detected cavity output field
is then obtained using the input-output relation $\hat{a}_{l}^{\mathrm{out}}(\omega)=\hat{a}_{l}^{\mathrm{in}}(\omega)-\sqrt{\kappa}\hat{a}_{l}(\omega))$
for the respective optical mode.

\subsection{1D QLT with amplitude or phase optical coupling\label{subsec:Single-mode-case-with}}

In this section we consider one mechanical mode, $\hat{b}_{j}$, and
one optical mode, $\hat{a}$, with two types of couplings: (i) $g_{j}(\hat{b}_{j}^{\dagger}+\hat{b}_{j})(\hat{a}^{\dagger}+\hat{a})$
and (ii) $g_{j}(\hat{b}_{j}^{\dagger}+\hat{b}_{j})i(\hat{a}^{\dagger}-\hat{a})$.
The former case (i) is the usual optomechanical coupling between the
mechanical mode and the amplitude quadrature of light which has been
reviewed in Sec.~\ref{soQLT}. To obtain the PSD one can restrict
the general multi-mode result in Eq.~\eqref{Scc} to the case of
one optical and one mechanical degree of freedom by setting $N=M=1$.
Alternatively, an explicit calculation of the PSD can be performed
by following the steps from Eqs.~\eqref{LAM}-\eqref{Nmatrix}. Specifically,
from Eq.~\eqref{matrix-standard} one first moves to the frequency
space and solves for the displacement operator $\hat{q}(\omega)=\hat{b}(\omega)+\hat{b}^{\dagger}(\omega)$,
i.e. the displacement operator is expressed in terms of noises $\hat{a}_{\text{in}}(\omega)$
and $\hat{b}_{\text{in}}(\omega)$. The PSD can then be readily obtained
by evaluating the expectation value using Eqs.~\eqref{kronecker}
and \eqref{Nmatrix}. The latter case (ii), where the mechanical mode
is now coupled to the phase quadrature of light, can be analysed using
analogous steps. Specifically, one first obtains the quantum Langevin
equations with the modified coupling and then follows the steps in
Eqs.~\eqref{LAM}-\eqref{Nmatrix}.

For both cases (i) and (ii) we can write the displacement operator
using the notation adopted in the main text: 
\begin{equation}
\hat{q}_{j}(\omega)\equiv\tilde{\mathcal{D}}_{j}^{1D}=M_{j}^{-1}\left[\sqrt{\Gamma}\tilde{Q}_{j}^{\text{therm}}+i\sqrt{\kappa}g_{j}\mu_{j}\tilde{Q}_{\Phi}^{\textrm{in}}\right],\label{1DDisplace1}
\end{equation}
where we have the normalization factor 
\begin{equation}
M_{j}(\omega)=1+g_{j}^{2}\mu_{j}(\omega)\eta^{(\Phi)}(\omega),
\end{equation}
the mechanical susceptibilities 
\begin{equation}
\mu_{j}(\omega)=\chi(\omega,\omega_{j},\Gamma)-\chi^{*}(-\omega,\omega_{j},\Gamma),\label{MSuscept}
\end{equation}
the mechanical noise 
\begin{equation}
\tilde{Q}_{j}^{\text{therm}}(\omega)=\chi(\omega,\omega_{j},\Gamma)\hat{b}_{j}^{\text{in}}(\omega)+\chi^{*}(-\omega,\omega_{j},\Gamma)\hat{b}_{j}^{\text{in}}{}^{\dagger}(\omega),
\end{equation}
the optical susceptibility 
\begin{equation}
\eta^{(\Phi)}(\omega)=e^{-i\Phi}\chi(\omega,-\Delta,\kappa)-e^{i\Phi}\chi^{*}(-\omega,-\Delta,\kappa),\label{OSuscept}
\end{equation}
the optical noise 
\begin{equation}
\tilde{Q}_{\Phi}^{\textrm{in}}(\omega)=e^{-i\Phi}\chi(\omega,-\Delta,\kappa)\hat{a}_{\text{in}}(\omega)+e^{i\Phi}\chi^{*}(-\omega,-\Delta,\kappa)\hat{a}_{\text{in}}^{\dagger}(\omega),
\end{equation}
and we have defined 
\begin{equation}
\chi(\omega,\omega_{j},\Gamma)=\left[-i(\omega-\omega_{j})+\frac{\Gamma}{2}\right]^{-1}.
\end{equation}

The above are (almost) the standard expressions for the 1D quantum
linear theory (QLT) of optomechanics. The only difference is that
we specify an angle $\Phi$ for the optical noise, such that $\Phi=0$
for standard optomechanical coupling, i.e. coordinates coupled to
the amplitude quadrature of light, but $\Phi=\pi/2$ for the coordinates
coupled to the phase quadrature of light (such as the $z$ coordinate
in the experiments in \cite{Delic2018}).

\subsection{3D QLT with amplitude and phase optical coupling\label{subsec:3D-QLT-with}}

In this section we consider three mechanical degrees of freedom that
are coupled to both the amplitude and phase quadratures of the optical
degree of freedom. Specifically, we consider the interaction Hamiltonian
given by: 
\begin{equation}
\frac{\hat{V}_{\text{int}}}{\hbar}=-\sum_{j}g_{jY}\hat{q}_{j}\hat{Y}-\sum_{j}g_{jP}\hat{q}_{j}\hat{P}-\sum_{j<k}g_{jk}\hat{q}_{j}\hat{q}_{k},\label{eq:3DQLT}
\end{equation}
where $\hat{Y}=\hat{a}^{\dagger}+\hat{a}$ and $\hat{P}=i(\hat{a}^{\dagger}-\hat{a})$,
and $\hat{q}_{j}=\hat{b}_{j}^{\dagger}+\hat{b}_{j}$ denotes the mechanical
degrees of freedom $\hat{x},\hat{y},\hat{z}$. The starting point
of the analysis are again the equations of motion written in frequency
domain: 
\begin{alignat}{1}
\hat{q}_{j}(\omega)= & J_{jY}(\omega)\hat{Y}(\omega)+J_{jP}(\omega)\hat{P}(\omega)+\sum_{k\neq j}J_{jk}(\omega)\hat{q}_{k}(\omega)+\sqrt{\Gamma}\tilde{Q}_{j}^{\text{therm}}(\omega),\label{eq:qj}\\
\hat{Y}(\omega)= & \sum_{j}J_{Yj}(\omega)\hat{q}_{j}(\omega)+\sqrt{\kappa}\tilde{Y}_{\text{in}}(\omega),\label{eq:Y}\\
\hat{P}(\omega)= & \sum_{j}J_{Pj}(\omega)\hat{q}_{j}(\omega)+\sqrt{\kappa}\tilde{P}_{\text{in}}(\omega),\label{eq:P}
\end{alignat}
where $J_{jk}(\omega)=ig_{jk}\mu_{j}(\omega)$ (for $j=x,y,z$, and
$k=x,y,z,Y,P$), $J_{Yj}(\omega)=i(\tilde{g}_{j}\chi(\omega,-\Delta,\kappa)-\tilde{g}_{j}^{*}\chi^{*}(-\omega,-\Delta,\kappa))$,
$J_{Pj}(\omega)=(\tilde{g}_{j}\chi(\omega,-\Delta,\kappa)+\tilde{g}_{j}^{*}\chi^{*}(-\omega,-\Delta,\kappa))$,
and we have defined the complex-valued couplings $\tilde{g}_{j}=g_{jY}+ig_{jP}$.
Note that in the main text, where we discuss a special case, we use
the more common notation of real-valued couplings: $g_{x}\equiv g_{xY}$,
and $g_{y}\equiv g_{yY}$, and $g_{z}\equiv g_{zP}$. The input noises
are given by 
\begin{alignat}{1}
\tilde{Q}_{j}^{\text{therm}}(\omega)= & \chi^{*}(-\omega,\omega_{j},\Gamma)b_{j}^{\text{in}\dagger}(\omega)+\chi(\omega,\omega_{j},\Gamma)b_{j}^{\text{in}}(\omega),\label{eq:Qth}\\
\tilde{Y}_{\text{in}}(\omega)= & \chi^{*}(-\omega,-\Delta,\kappa)a_{\text{in}}^{\dagger}(\omega)+\chi(\omega,-\Delta,\kappa)a_{\text{in}}(\omega),\\
\tilde{P}_{\text{in}}(\omega)= & i(\chi^{*}(-\omega,-\Delta,\kappa)a_{\text{in}}^{\dagger}(\omega)-\chi(\omega,-\Delta,\kappa)a_{\text{in}}(\omega)),\label{eq:Pin}
\end{alignat}
where $\tilde{Q}_{j}^{\text{therm}}(\omega)$ denotes the mechanical
noises $\tilde{X}^{\text{therm}}(\omega)$, $\tilde{Y}^{\text{therm}}(\omega)$,
$\tilde{Z}^{\text{therm}}(\omega)$. 

It is instructive to separate the contributions to the spectra of
$\hat{q}_{j}(\omega)$ into two categories: one that contains the
terms of a 1D approximation and one that contains additional terms
arising in a realistic 3D problem. Specifically, using Eqs.~\eqref{eq:Y}
and \eqref{eq:P} we can rewrite Eq.~\eqref{eq:qj} as:

\begin{equation}
\hat{q}_{j}(\omega)=\tilde{\mathcal{D}}_{j}^{1D}+\sum_{k\neq j}\mathcal{G}_{jk}^{3D}(\omega)\hat{q}_{k}(\omega),\label{3DqNoise}
\end{equation}
where $\tilde{\mathcal{D}}_{j}^{1D}$ is the displacement noise already
present in 1D problems, and $\mathcal{G}_{jk}^{3D}$ are new 3D couplings.
In the main text we made the low order approximation $\hat{q}_{k}(\omega)\to\tilde{\mathcal{D}}_{k}^{1D}$
to allow a simple analysis. 

For example, for the special case of 3D coherent scattering discussed
in the main text we find (see Sec.~\ref{sec:-3D-levitated} for more
details): 

\begin{alignat}{1}
\mathcal{G}_{jk}^{3D}(\omega) & =\frac{i\mu_{j}(\omega)}{M_{j}(\omega)}\left[i\eta^{(0)}(\omega)g_{j}g_{k}+g_{jk}\right],\label{eq:jk}\\
\mathcal{G}_{jz}^{3D}(\omega) & =\frac{i\mu_{j}(\omega)}{M_{j}(\omega)}\left[-i\eta^{(\pi/2)}(\omega)g_{j}g_{z}+g_{jz}\right],\\
\mathcal{G}_{zj}^{3D}(\omega) & =\frac{i\mu_{z}(\omega)}{M_{z}(\omega)}\left[i\eta^{(\pi/2)}(\omega)g_{z}g_{j}+g_{jz})\right],\label{zj}
\end{alignat}
where in Eqs.~\eqref{eq:jk}-\eqref{zj} the indices $j,k$ denote
$x$ or $y$. 

We now continue with the general analysis. For numerical accuracy,
we here give the exact expressions for the displacements in terms
of noises. Specifically, starting from Eqs.~(\ref{eq:qj})-(\ref{eq:P})
we eventually find:

\begin{alignat}{1}
\hat{q}_{j} & (\omega)=A_{j}(\omega)\tilde{Y}_{\text{in}}(\omega)+B_{j}(\omega)\tilde{P}_{\text{in}}(\omega)+C_{j}(\omega)\hat{X}^{\text{therm}}(\omega)+D_{j}(\omega)\tilde{Y}^{\text{therm}}(\omega)+D_{j}(\omega)\tilde{Z}^{\text{therm}}(\omega),\label{eq:xs}
\end{alignat}
where $j$ denotes one of the mechanical motions, 
\begin{alignat}{1}
A_{j} & =N(\xi_{jx}\beta_{xY}+\xi_{jy}\beta_{yY}+\xi_{jz}\beta_{zY}),\\
B_{j} & =N(\xi_{jx}\beta_{xP}+\xi_{jy}\beta_{yP}+\xi_{jz}\beta_{zP}),\\
C_{j} & =N\xi_{jx}\beta_{xx},\\
D_{j} & =N\xi_{jy}\beta_{yy},\\
E_{j} & =N\xi_{jz}\beta_{zz},
\end{alignat}
where $\beta_{jY}=N_{j}J_{jY}$, $\beta_{jP}=N_{j}J_{jP}$, $\beta_{jj}=N_{j}$,
$N_{j}=(1-J_{jY}J_{Yj}-J_{jP}J_{Pj})^{-1}$. We have defined the coefficients
$\xi_{jj}=1-\frac{1}{2}R_{kl}R_{lk}$ (with $l,k\neq j$ and $k\neq l$),
$\xi_{jk}=R_{jk}+R_{jl}R_{lk}$(with $j\neq k$, $l\neq k$, and $l\neq j$),
and $R_{jk}=N_{j}(J_{jY}J_{Yk}+J_{jP}J_{Pk}+J_{jk})$. The overall
normalization is given by $N=(1-\frac{1}{2}\sum R_{kl}R_{lk}-\frac{1}{3}\sum R_{kl}R_{lj}R_{jk})^{-1}$
(with $l\neq k$,$j\neq k$, and $l\neq j$). The PSDs can be readily
obtained from Eq.~\eqref{eq:xs} using the methods discussed in Sec.~\ref{soQLT}.

\subsection{Self-energy, optical spring, and damping}
In this section we obtain the analytical expressions for the  self-energy, relevant to the experiments 
of \cite{Delic2019} that couple $x$ and $y$ (but not significantly $z$) . We obtain also the resulting optical spring and damping formulae. We start from the coupled equations:
\begin{alignat}{1}
\hat{x} & =J_{xy}\hat{Y}+\tilde{Q}_{x}^{\text{therm}},\\
\hat{y} & =J_{yx}\hat{Y}+\tilde{Q}_{y}^{\text{therm}},\\
\hat{Y} & =J_{Yx}\hat{x}+J_{Yy}\hat{y}+\tilde{Y}_{\text{in}}.
\end{alignat}
We now focus on the $x$-motion, while the formulare for the $y$ motion can be obtained by formally exchanging $x\longleftrightarrow y$ in the formulae. Specifically, we solve for $\hat{x}$ to find:
\begin{equation}
\hat{x}=J_{xY}\left[1-J_{Yy}J_{yY}\right]^{-1}\left(J_{Yx}\hat{x}+J_{Yy}\tilde{Q}_{y}^{\text{therm}}+\tilde{Y}_{\text{in}}\right)+\tilde{Q}_{x}^{\text{therm}}.
\end{equation}
One can then extract the self-energy $\Sigma_{x}$, which is given
by:

\begin{equation}
\mu_{x}\Sigma_{x}\equiv-\frac{J_{xY}J_{Yx}}{1-J_{Yy}J_{yY}}.\label{eq:selfenergy}
\end{equation}
We note that the numerator is the usual term $\propto g_{x}^{2}$
which arises already in the 1D analysis, while the denominator term
$\propto g_{y}^{2}$ is a new effect which arises in the 3D analysis.
In particular, considering
the expressions for $J_{xY}$ and $J_{Yx}$ we find from Eq.~(\ref{eq:selfenergy}):

\begin{equation}
\Sigma_{x}\equiv\frac{g_{x}^{2}\eta^{(0)}}{1+g_{y}^{2}\mu_{y}\eta^{(0)}}.
\end{equation}
Finally, we can find the change of the damping, $\delta\Gamma$, and
the shift of frequency, $\delta\omega$, using the following expressions~\cite{marquardt2007quantum}:
\begin{equation}
\delta\Gamma_{j}=\frac{\text{Im}(\Sigma_{j}(\omega_{\text{j}}))}{\omega_{j}},\qquad\delta\omega_{j}=\frac{\text{Re}(\Sigma_{j}(\omega_{\text{j}}))}{2\omega_{\text{j}}},\label{eq:change}
\end{equation}
where $\omega_{\text{j}}$ denotes the mechanical frequency.

\section{3D levitated optomechanics in a cavity driven by coherently scattered
tweezer light}\label{sec:-3D-levitated}

We consider the 3D optomechanical system such as the coherent scattering
cavity levitation introduced in \cite{Windey2018,Delic2018,GonzalezBallestero2019}.
As discussed below, some of the optomechanical coupling terms are
of the form $ig_{k}({\hat{a}}^{\dagger}-{\hat{a}})({\hat{b}}_{k}^{\dagger}+{\hat{b}}_{k})$,
i.e. the mechanical motion can couple to the phase quadrature of the
light, in addition to the more typical coupling to the amplitude quadrature,
i.e. $g_{k}({\hat{a}}^{\dagger}+{\hat{a}})({\hat{b}}_{k}^{\dagger}+{\hat{b}}_{k})$.
Specifically, we will consider the case when the $z$ motion has the
former type, while $x$ and $y$ motions have the latter one.

In addition, for a truly 3D system, one allows also direct couplings
between the mechanical modes, i.e. $g_{kk'}\hat{q}_{k}\hat{q}_{k'}$,
where $k,k'\equiv x,y,z$. Direct couplings are not usually considered
in optomechanics: although multi-mode systems are commonly studied
(such as multiple vibration modes of membranes) coupling between mechanical
modes is not usually of interest. However for the considered 3D optical
levitation this is not only important, but the $g_{kk'}$ are closely
correlated with the couplings $g_{k}$. Specifically, we will find
$g_{kk'}\propto g_{k}g_{k'}$, which has important consequences for
sensing. 

In particular, in a cavity populated only by scattered light as in
the recent 3D set-ups in levitated optomechanics, we need consider
only a single light mode, but three mechanical modes including direct
coupling: 
\begin{equation}
\hat{H}=\hat{H}_{0}-(\hat{a}^{\dagger}+\hat{a})[g_{x}(\hat{b}_{x}^{\dagger}+\hat{b}_{x})+g_{y}(\hat{b}_{y}^{\dagger}+\hat{b}_{y})]-i(\hat{a}^{\dagger}-\hat{a})g_{z}(\hat{b}_{z}^{\dagger}+\hat{b}_{z})-\sum_{k}\sum_{j\neq k}g_{jk}(\hat{b}_{k}^{\dagger}+\hat{b}_{k})(\hat{b}_{j}^{\dagger}+\hat{b}_{j})\label{H3D}
\end{equation}
where ${\hat{H}}_{0}=-\Delta{\hat{a}}^{\dagger}{\hat{a}}+\sum_{k=x,y,z}\omega_{k}{\hat{b}}_{k}^{\dagger}{\hat{b}}_{k}$
(see Sec.~\ref{subsec:3D-QLT-with} where we have developed a generic
framework to solve such Hamiltonians within QLT). In order to extract
the dynamical parameters, i.e. the frequencies $\omega_{k}$, the
optomechanical couplings $g_{k}$, and the direct couplings $g_{kj}$,
we must first consider the physical tweezer and cavity potentials
(Sec.~\ref{subsec:3D-coherent-scattering}), and then expand to quadratic
order around an equilibrium position (Sec.~\ref{subsec:Quadratic-Hamiltonian:-unified}). 

\subsection{3D coherent scattering Hamiltonian\label{subsec:3D-coherent-scattering}}

We consider the hybrid tweezer-cavity experiments introduced in~\cite{Delic2018},
which employed set-ups very similar to those in~\cite{Windey2018,GonzalezBallestero2019}.
A nanoparticle is trapped at the focus of a tweezer field and interacts
with light coherently scattered from the tweezer field into the (undriven)
cavity.

The Hamiltonian describing the interaction between the nanoparticle
and the combined fields of the tweezer and cavity is given by:

\begin{equation}
\hat{H}=-\frac{\alpha}{2}\vert\mathbf{\hat{E}}_{\text{cav}}+\mathbf{\hat{E}}{}_{\text{tw}}\vert^{2},\label{eq:Hch}
\end{equation}
where $\mathbf{\hat{E}}_{\text{cav}}$ ($\mathbf{\hat{E}}_{\text{tw}}$)
denotes the cavity (tweezer) field, $\alpha=3\epsilon_{0}V_{s}\frac{\epsilon_{R}-1}{\epsilon_{R}+2}$
is the polarizability of the nanosphere, $V_{s}$ is the volume of
the nanosphere, $\epsilon_{0}$ is the permittivity of free space,
and $\epsilon_{R}$ is the relative dielectric permittivity.

We assume a coherent Gaussian tweezer field and replace the modes
with c-numbers to find:

\begin{equation}
\mathbf{\mathbf{\hat{E}}}_{\text{tw}}=\frac{\epsilon_{tw}}{2}\frac{1}{\sqrt{1+(\frac{z}{z_{R}})^{2}}}e^{-\frac{\hat{x}^{2}}{w_{x}^{2}}}e^{-\frac{\hat{y}^{2}}{w_{y}^{2}}}e^{ik\hat{z}+i\Phi(\hat{z})}e^{-i\omega_{\text{tw}}t}\mathbf{e}_{y}+\text{c.c.},\label{eq:Etw}
\end{equation}
where $\Phi(z)=-\arctan\frac{z}{z_{R}}$ is the Gouy phase, $z_{R}=\frac{\pi w_{x}w_{y}}{\lambda}$
is the Rayleigh range, $w_{x}$ ($w_{y}$) are the beam waist along
the $x$ ($y$) axis, $\epsilon_{tw}=\sqrt{\frac{4P_{\text{tw}}}{w_{x}w_{y}\pi\epsilon_{0}c}}$
is the amplitude of the electric field, $c$ is the speed of light,
$P_{\text{tw}}$ is the laser power, $\omega_{\text{tw}}$ is the
tweezer angular frequency, $t$ is the time, and $\hat{\boldsymbol{r}}=(\hat{x},\hat{y},\hat{z})$
is the position of the nanoparticle. $\mathbf{e}_{j}$ are the unit
vectors: $\mathbf{e}_{z}$ is aligned with the symmetry axis of the
tweezer field and $\mathbf{e}_{y}$ is aligned with the polarization
of the tweezer field. 

The cavity field is given by:

\begin{equation}
\mathbf{\hat{E}}_{\text{cav}}=\epsilon_{c}\text{cos}(k(x_{0}^{\text{(c)}}+\hat{x}^{\text{(\text{c)}}}))\mathbf{e}_{y}^{c}\left[\hat{a}+\hat{a}^{\dagger}\right],\label{eq:Ecav}
\end{equation}
where $\epsilon_{c}=\sqrt{\frac{\hbar\omega_{c}}{2\epsilon_{0}V_{c}}}$
is the amplitude at the center of the cavity, $V_{c}$ is the cavity
volume, $\omega_{c}$ is the cavity frequency, $\hat{a}$ ($\hat{a}^{\dagger}$
)is the annihilation (creation) operator, $x_{0}^{\text{(c)}}$ is
an offset of the cavity coordinate system (centered at a cavity antinode)
with respect to the tweezer coordinate system. The cavity $x_{c}$-$y_{x}$
plane is rotated by an angle $\theta$ with respect to the tweezer
$x$-$y$ plane:

\begin{alignat}{1}
\left[\begin{array}{c}
x_{c}\\
y_{c}
\end{array}\right] & =\left[\begin{array}{cc}
\text{sin}(\theta) & \text{cos}(\theta)\\
-\text{cos}(\theta) & \text{sin}(\theta)
\end{array}\right]\left[\begin{array}{c}
x\\
y
\end{array}\right].
\end{alignat}
Note that for $\theta=0$ the tweezer polarization ($y$-axis) becomes
aligned with the cavity symmetry axis ($x_{c}$-axis). In particular,
we have $\hat{x}^{\text{(\text{c)}}}=\text{sin}(\theta)\hat{x}+\text{cos}(\theta)\hat{y}$.
Furthermore, we then have the following relation between the cavity
and tweezer unit vectors

\begin{alignat}{1}
\mathbf{e}_{y}^{c} & =\left[-\mathbf{e}_{x}\text{cos}(\theta)+\mathbf{e}_{y}\text{sin}(\theta)\right].\label{eq:eyc}
\end{alignat}
We expand the Hamiltonian in Eq.~(\ref{eq:Hch}) exploiting Eqs.~(\ref{eq:Etw}),
(\ref{eq:Ecav}), and (\ref{eq:eyc}) to obtain three terms:

\begin{equation}
\hat{H}=-\frac{\alpha}{2}\vert\mathbf{\mathbf{\hat{E}}}_{\text{tw}}\vert^{2}-\frac{\alpha}{2}\vert\mathbf{\hat{E}}_{\text{cav}}\vert^{2}-\frac{\alpha\text{sin}(\theta)}{2}(\mathbf{\hat{E}}_{\text{cav}}^{\dagger}\mathbf{\mathbf{\hat{E}}}_{\text{tw}}+\mathbf{\hat{E}}_{\text{cav}}\mathbf{\mathbf{\hat{E}}}_{\text{tw}}^{\dagger}),\label{eq:Hch2}
\end{equation}
where the terms on the right hand-side are the tweezer term, the cavity
term, and the tweezer-cavity interaction term (from left to right).
The first (tweezer field) term dominates the trapping and primarily
sets the three mechanical frequencies $\omega_{x}$, $\omega_{y}$,
and $\omega_{z}$. The second term provides a (typically) small correction
to the frequencies and is included only for numerical precision. The
third term, which we will denote as $\hat{V}_{\text{int}}$, is the
most interesting and novel form of optomechanical interaction. As
discussed in \cite{Delic2018,GonzalezBallestero2019}, time-dependencies
in this term are eliminated through rotating frame approximations
leaving an effective optomechanical Hamiltonian: 
\begin{equation}
\frac{\hat{V}_{\text{int}}}{\hbar}=-E_{d}\text{cos}(\phi+k(\hat{x}\sin\theta+\hat{y}\cos\theta)\left[\hat{a}e^{-i(k\hat{z}+\Phi(\hat{z}))}+\hat{a}^{\dagger}e^{+i(k\hat{z}+\Phi(\hat{z}))}\right],\label{eq:Hopto}
\end{equation}
where $E_{d}=\frac{\alpha\epsilon_{c}\epsilon_{tw}\sin\theta}{2\hbar}$,
$\phi=kx_{0}^{\text{(c)}}$ represents the effect of the shift between
the origin of the cavity and tweezer. The experiments allow positioning
of $x_{0}^{\text{(c)}}$ with an accuracy of $\sim8$ nm for $\lambda=1064$nm.
In first approximation one can neglect the Gouy phase $\Phi$. Linearisation
of the above Hamiltonian to quadratic order yields the 3D optomechanical
couplings.

\subsection{Quadratic Hamiltonian: unified form\label{subsec:Quadratic-Hamiltonian:-unified}}

We expand the Hamiltonian in Eq.~(\ref{eq:Hopto}) around an equilibrium
point $(x_{0},y_{0},z_{0},{\bar{\alpha}})^{\top}$ by making the substitution
$(x,y,z,a)^{\top}\rightarrow(x_{0},y_{0},z_{0},{\bar{\alpha}})^{\top}+(x,y,z,a)^{\top}$,
where $(x,y,z,a)^{\top}$ on the right hand-side denotes small fluctuations. 

To a first approximation, $x_{0},y_{0},z_{0}$ represents the origin
of the strong tweezer trap; however, as investigated in \cite{GonzalezBallestero2019},
when the cavity is strongly populated, this must be corrected with
further small offsets $\delta x_{0},\delta y_{0},\delta z_{0}$. These
emerge naturally from numerical simulations and can also be well estimated
through the linearisation analysis. The mean cavity photon occupancy
number $n=|{\bar{\alpha}}|^{2}$ may also need to be corrected from
the approximate form \cite{Delic2018} ${\bar{\alpha}}=-iE_{d}\cos(\phi)/(i\Delta-\kappa/2)$
to allow for the fact that $\phi\simeq kx_{0}^{\text{(c)}}+\delta x_{0}\cos\theta+\delta y_{0}\sin\theta$
must include the additional corrections to $x_{0}^{\text{(c)}}$.

A unique feature of these new levitated set-ups is that the optomechanical
coupling can be via the momentum quadrature. In the calculation in~\cite{Delic2018,Windey2018,GonzalezBallestero2019}
this affected only the $z$ coordinate. However, we note that if there
are significant offsets in the fields or misalignment of the cavity
and tweezer axes, in general, one might wish to consider both amplitude
and momentum couplings to all mechanical modes so here we introduce
a unified notation.

It is convenient to introduce the notation $\hat{\mathsf{Y}}=\hat{a}^{\dagger}+\hat{a}$,
and $\mathsf{\hat{P}}=i\left(\hat{a}^{\dagger}-\hat{a}\right)$ for
the optical field. We also similarly use $\hat{x}=x_{\text{zpf}}\left(\hat{b}_{x}^{\dagger}+\hat{b}_{x}\right)$,
$y=y_{\text{zpf}}\left(\hat{b}_{y}^{\dagger}+\hat{b}_{y}\right)$,
$z=z_{\text{zpf}}\left(\hat{b}_{z}^{\dagger}+\hat{b}_{z}\right)$,
where zero-point fluctuation lengths are given by $x_{\text{zpf}}=\sqrt{\frac{\hbar}{2m\omega_{x}}}$,
$y_{\text{zpf}}=\sqrt{\frac{\hbar}{2m\omega_{y}}},$ and $z_{\text{zpf}}=\sqrt{\frac{\hbar}{2m\omega_{z}}}$,
and $m$ is the mass of the levitated nanoparticle.

Redefining $\hat{x}/x_{\text{zpf}}\rightarrow\hat{x}$, $\hat{y}/y_{\text{zpf}}\rightarrow\hat{y}$,
and $\hat{z}/z_{\text{zpf}}\rightarrow\hat{z}$ we write:

\begin{equation}
\frac{\hat{H}}{\hbar}=-\left[g_{xy}\hat{x}\hat{y}+g_{xz}\hat{x}\hat{z}+g_{yz}\hat{y}\hat{z}+(g_{xY}\hat{x}+g_{yY}\hat{y}+g_{zY}\hat{z})\hat{Y}+(g_{xP}x+g_{yP}y+g_{zP}z)\hat{P}\right],
\end{equation}
where we have omitted the harmonic oscillator terms. We can also rewrite
the optical quadratures in terms of the mode operator $\hat{a}$:

\begin{equation}
\frac{\hat{H}}{\hbar}=-\left[g_{xy}xy+g_{xz}z+g_{yz}yz+(\tilde{g}_{x}x+\tilde{g}_{y}y+\tilde{g}_{z}z)a^{\dagger}+(\tilde{g}_{x}^{*}x+\tilde{g}_{y}^{*}y+\tilde{g}_{z}^{*}z)a\right],
\end{equation}
where we have introduced the complex-valued couplings $\tilde{g}_{j}=g_{jY}+ig_{jP}$
(see Sec.~\ref{subsec:3D-QLT-with} for the resolution of this general
Hamiltonian within 3D QLT). However in the following we opt to use
the more conventional notation introduced for the special case in
Eq.~\eqref{H3D} where all the coupling constants are defined as
real-valued. Specifically, from Eq.~\eqref{eq:Hopto} we find the
following non-zero light-matter couplings:

\begin{alignat}{1}
g_{x}\equiv g_{xY}= & -E_{d}k\sin(\text{\text{\text{\ensuremath{\theta}}}})\sin(\phi)x_{\text{zpf}},\\
g_{y}\equiv g_{yY}= & -E_{d}k\cos(\text{\text{\text{\ensuremath{\theta}}}})\sin(\phi)y_{\text{zpf}},\\
g_{z}\equiv g_{zP}= & E_{d}k\cos(\phi)z_{\text{zpf}}.
\end{alignat}
In addition we also have matter-matter couplings

\begin{alignat}{1}
g_{xy}= & -E_{d}k^{2}Y_{0}\sin(\text{\ensuremath{\theta}})\cos(\text{\ensuremath{\theta}})\cos(\phi)x_{\text{zpf}}y_{\text{zpf}},\\
g_{xz}= & -E_{d}k^{2}P_{0}\sin(\text{\ensuremath{\theta}})\sin(\phi)x_{\text{zpf}}z_{\text{zpf}},\\
g_{yz}= & -E_{d}k^{2}P_{0}\cos(\text{\ensuremath{\theta}})\sin(\phi)y_{\text{zpf}}z_{\text{zpf}}.
\end{alignat}
The harmonic frequencies are given by 

\begin{alignat}{1}
\omega_{j} & =\sqrt{\frac{1}{m}(T_{j}+C_{j}+T_{j}^{c})},
\end{alignat}
where $T_{x}=\frac{\alpha\epsilon_{tw}^{2}}{w_{x}^{2}}$,$T_{y}=\frac{\alpha\epsilon_{tw}^{2}}{w_{y}^{2}}$,
and $T_{z}=\frac{\alpha\epsilon_{tw}^{2}}{2z_{R}^{2}}$ are the typically
dominant contributions arising from the tweezer trap. The corrections
from the cavity are $C_{x}=2\alpha\epsilon_{c}^{2}k^{2}n\sin^{2}(\text{\ensuremath{\theta}})\text{cos}(2\phi)$
and $C_{y}=2\alpha\epsilon_{c}^{2}k^{2}n\text{cos}^{2}(\text{\ensuremath{\theta}})\text{cos}(2\phi)$,
and $C_{z}=0$. The contributions arising from the coupling between
the cavity and tweezer are $T_{x}^{c}=\frac{\hbar E_{d}}{w_{x}^{2}}(2+k^{2}w_{x}^{2}\text{sin}^{2}(\theta))Y_{0}\text{cos}(\phi)$,
$T_{y}^{c}=\frac{\hbar E_{d}}{w_{y}^{2}}(2+k^{2}w_{y}^{2}\text{cos}^{2}(\theta))Y_{0}\text{cos}(\phi)$,
and $T_{z}^{c}=\frac{\hbar E_{d}}{z_{R}^{2}}(1+k^{2}z_{R}^{2})Y_{0}\text{cos}(\phi)$.
We remark that corrections from $C_{j}$ and $T_{j}^{c}$ can in certain
cases become important, e.g. when the cavity has a high photon occupancy,
potentially even leading to nanoparticle loss. The cavity-tweezer
interaction also changes the cavity detuning from $\Delta$ to $\Delta+\Delta_{0}$,
where $\Delta_{0}=\frac{\alpha\epsilon_{c}^{2}}{\hbar}\cos^{2}(\phi)$.

% \bibliographystyle{unsrt}
%\bibliography{Opto}
\putbib
\end{bibunit}
%\end{widetext}
\end{document}